\pdfoutput=1
\documentclass[aps,twocolumn,prd,superscriptaddress,showpacs,preprintnumbers,
amsmath,amssymb,nofootinbib]{revtex4}

\usepackage{amsmath,amssymb}
\usepackage{graphicx}
\usepackage{rotating}
\usepackage{bm}
\usepackage[usenames]{color}
\usepackage[colorlinks=true,linkcolor=blue,citecolor=blue,urlcolor=blue]{
hyperref}
\usepackage{epsfig}
\usepackage{array} 
 
\newcommand{\tr}{\ensuremath{\operatorname{tr}}}

\def\eq#1{Eq.~(\ref{#1})}
\def\eqs#1{Eqs.~(\ref{#1})}
\def\Fig#1{Fig.~\ref{#1}}

\newcommand{\Phibar}{\ensuremath{\bar{\Phi}}}

\def\dr{{D\!\llap{/}}\,}

\def\0#1#2{\frac{#1}{#2}}

\begin{document}
\title{On the phase structure and thermodynamics of QCD}
\author{Tina Katharina Herbst}
\affiliation{Institut f\"{u}r Physik, Karl-Franzens-Universit\"{a}t,
  A-8010 Graz, Austria}
\affiliation{Institut f\"ur Theoretische Physik,
University of Heidelberg, 
Philosophenweg 16, D-69120 Heidelberg, Germany}
\author{Jan M. Pawlowski}
\affiliation{Institut f\"ur Theoretische Physik, University of Heidelberg, 
Philosophenweg 16, D-69120 Heidelberg, Germany}
\affiliation{ExtreMe Matter Institute EMMI, GSI, Planckstr. 1, D-64291
Darmstadt, Germany.}
\author{Bernd-Jochen Schaefer}
\affiliation{Institut f\"{u}r Physik, Karl-Franzens-Universit\"{a}t,
  A-8010 Graz, Austria}

\pacs{12.38.Aw, 
11.10.-z	, 
11.30.Rd	, 
12.38.-t}		

\begin{abstract} 
  We discuss the phase structure and thermodynamics of QCD by means of
  dynamical chiral effective models. Quark and meson fluctuations are
  included via the functional renormalization group. We study the
  influence of confinement in addition to the impact of fluctuations
  by comparing the results of the chiral models to their Polyakov-loop
  extended versions. Furthermore, we discuss the mass sensitivity of
  the phase structure and thermodynamics and find interesting
  modifications close to the chiral limit.
\end{abstract}

\maketitle

\section{Introduction}
\label{sec:intro}

The properties of strongly-interacting matter at nonvanishing
temperature and density are in the focus of many theoretical and
experimental efforts. On the experimental side, future and running
heavy-ion experiments at various facilities such as GSI, JINR, CERN
and BNL aim at probing the phase structure of quantum chromodynamics
(QCD), especially in the regime where potentially a critical endpoint
(CEP) is present.

On the theoretical side, functional continuum methods are well suited
for a combined study of the confinement-deconfinement and chiral phase
structure of QCD at finite density. In recent years, functional
renormalization group (FRG) studies and Dyson-Schwinger (DSE) studies have
provided valuable insights into the phase structure of strongly
interacting matter. These studies were performed both within first
principle QCD as well as within low-energy effective models,
see~\cite{Braun:2009gm, Skokov:2010wb,Herbst:2010rf, Pawlowski:2010ht,
  Braun:2011fw, Braun:2012zq,Fischer:2012vc}. For QCD-related reviews
see~\cite{Litim:1998nf, Berges:2000ew, Polonyi:2001se, Pawlowski:2005xe,
Gies:2006wv, Schaefer:2006sr,Pawlowski:2010ht,Schaefer:2011pn, Braun:2011pp,
vonSmekal:2012vx} (FRG) and \cite{Alkofer:2000wg, Roberts:2000aa,
Fischer:2006ub, Fischer:2008uz, Binosi:2009qm,Maas:2011se} (DSE).

At vanishing density, lattice simulations provide additional
insights from first principle QCD that are complementary to the
continuum studies. This also helps to improve the systematic error analysis
of the respective approaches. At nonvanishing chemical potential, however,
lattice simulations are hampered by the sign problem, see
e.g.~\cite{Karsch:2001cy, Philipsen:2007rj}. Impressive progress in
overcoming this problem has been made~\cite{deForcrand:2010ys,
  Cea:2009ba, Ejiri:2012ng, Aarts:2008wh, Schmidt:2006us,
  Seiler:2012wz,Aarts:2012ft}, but to date lattice QCD is still
restricted to small chemical potential.

Low-energy effective models, such as the Nambu--Jona-Lasinio (NJL) and
quark-meson (QM) models, are known to describe the chiral dynamics of
QCD rather well; for a recent review see e.g. \cite{Fukushima:2013rx}. 
Polyakov-loop extended versions thereof (PNJL, PQM)~\cite{Megias:2004hj,
Fukushima:2003fw, Schaefer:2007pw}, have advanced our understanding of the
confinement-deconfinement aspects of the QCD phase structure, 
e.g.~\cite{Meisinger:1995ih, Pisarski:2000eq, Fukushima:2003fw, Mocsy:2003qw,
Dumitru:2003hp, Ratti:2005jh, Ghosh:2006qh, Sasaki:2006ww, Schaefer:2007pw,
Sakai:2008py, Schaefer:2009ui, Skokov:2010wb, Herbst:2010rf, Mintz:2012mz}.  It
has been known for a long time, e.g.~\cite{Berges:2000ew,Pawlowski:2005xe,
  Gies:2006wv,Braun:2011pp,Schaefer:2011pn}, that low-energy effective
models can be systematically related to full QCD within the FRG approach. For
Polyakov-loop extended models this follows from the Landau gauge approach in
\cite{Braun:2009gm, Haas:2010bw, Pawlowski:2010ht}, and for the Polyakov gauge,
see \cite{Marhauser:2008fz,Kondo:2010ts}. This setting has been discussed in
detail for the PQM model in \cite{Pawlowski:2010ht,Herbst:2010rf,
Herbst:2012ht}; see also our discussion in Sec.~\ref{sec:PQM}.  While chiral
symmetry and its dynamical breaking are well described within these models,
confinement is only included in a statistical manner. Moreover, the glue
potential of full QCD, encoding the gauge dynamics in the presence of matter
fields, is replaced by a Polyakov-loop potential. This potential is
fixed to lattice data of the pure Yang-Mills system at vanishing chemical
potential. In such an approach, the coupling of the matter sector to the gauge
sector is lost, e.g. \cite{Schaefer:2007pw,  Herbst:2010rf}. This is discussed
in detail below, where we show how the matter back-coupling can effectively be
taken into account.

In conclusion, the FRG approach to the low-energy dynamics of QCD is a
promising setup for investigating the QCD phase structure and
thermodynamics. A study along these lines has already been put forward
in~\cite{Herbst:2010rf, Herbst:2012ht}. In the present work we extend
this approach by a more refined inclusion of the matter and glue dynamics 
as well as by studying the mass sensitivity of various observables. 

This paper is structured as follows: in Sec.~\ref{sec:PQM} we introduce the
Polyakov--quark-meson model as a low-energy truncation of full two-flavor QCD.
Furthermore, the inclusion of the matter back-coupling to the gauge sector is
discussed. Section~\ref{subsec:flow} summarizes the renormalization group
approach and the resulting flow equation for our model. In the following
Sec.~\ref{sec:phase} we compare the phase structure of the PQM model to that of
the QM model and discuss the mass sensitivity of the phase transitions. In
particular, we find an intriguing splitting of the chiral phase transition at
low Goldstone-boson masses, which is only observed when fluctuations are taken
into account. Some thermodynamic observables and their mass dependence are
studied in Sec.~\ref{sec:TD}. Concluding remarks and an outlook are given in
Sec.~\ref{sec:conclusion} and some details of our numerical implementation can
be found in the Appendix.

\section{From QCD to Polyakov-loop extended chiral models} 
\label{sec:PQM}

We have already emphasized in the introduction that low-energy effective models
can be related to full QCD in a systematic fashion. 
In \cite{Braun:2009gm, Pawlowski:2010ht, Marhauser:2008fz, Kondo:2010ts,
Herbst:2010rf, Herbst:2012ht} this is discussed for the embedding of Polyakov
loop extended models: The FRG provides a setting which allows to approach
the low temperature regime of QCD from large momentum scales by successively
integrating out momentum shells. This entails that one starts at asymptotically
large momenta with perturbative QCD with one coupling parameter, the strong
coupling $\alpha_s(p^2)$. The other parameters are the entries in the quark
mass matrix. In the present two-flavor study the masses of up and down quark are
small and the mass matrix is diagonal in flavor space.

\begin{figure}[t]
  \includegraphics[width=\columnwidth]{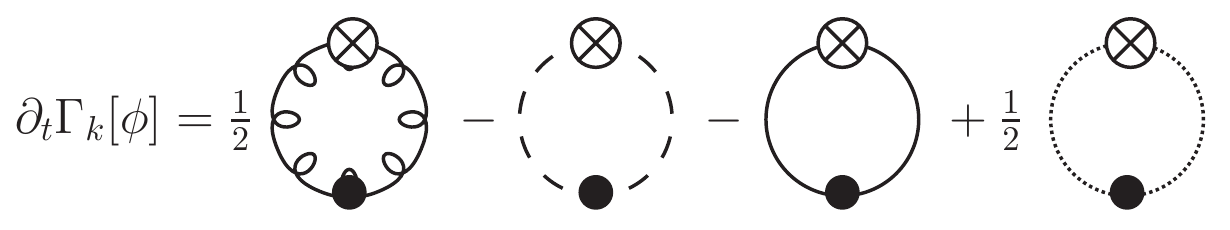}
  \caption{Partially hadronized version of the FRG flow for QCD. The
    loops denote the gluon, ghost, quark and meson contributions,
    respectively. The crosses mark the RG regulator term.}
  \label{fig:QCDflow}
\end{figure}

\begin{figure}[t]
  \includegraphics[width=.55\columnwidth]{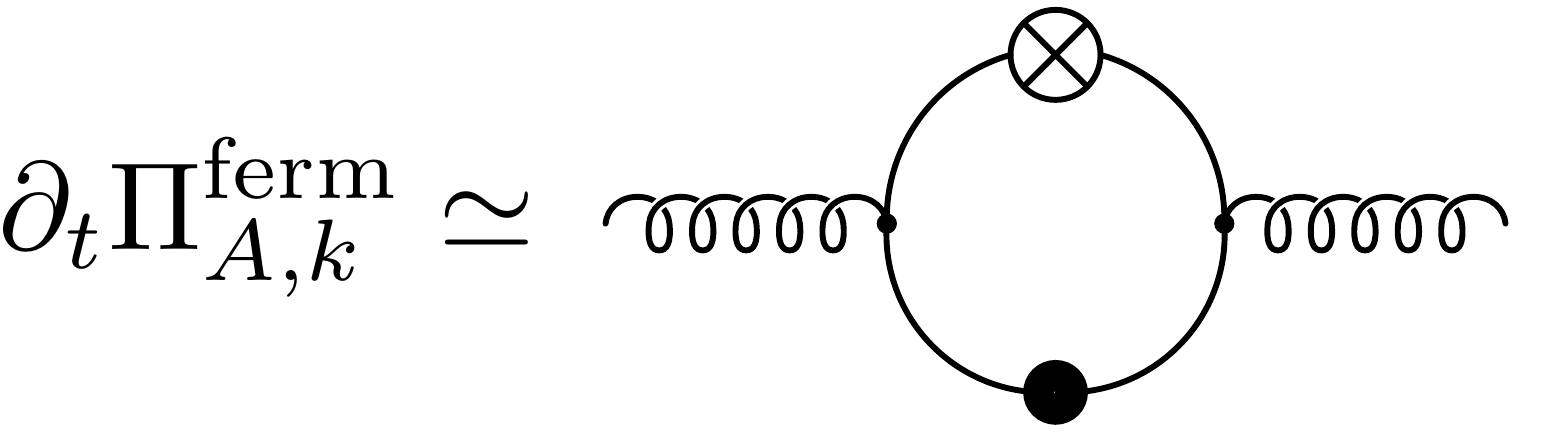}
  \caption{Quark polarization contribution to the gluon propagator
  representing a contribution to the matter back-coupling.}
  \label{fig:quark_loop}
\end{figure}

When lowering the momentum scale one systematically includes quark and gluon
fluctuations into the theory, finally approaching the hadronic phase; see
\cite{Braun:2008pi} for one flavor QCD and \cite{Braun:2009gm,Pawlowski:2010ht}
for two-flavor QCD. Close to the phase boundary between the quark-gluon plasma
phase and the hadronic phase and at not too large chemical potential, mesonic
degrees of freedom, in particular the pion and sigma fluctuations, become
important. The related quark bilinears, $\bar q q$ and $\bar q \gamma_5\vec\tau
q$, carry the same quantum numbers, i.e.  they have a considerable overlap with
the full meson operators. The matter sector of the QCD effective action can be
conveniently written in powers of these bilinears and related kinetic terms. The
strength and the momentum dependence is systematically computed by means of the
flow equation for QCD, depicted in Fig.~\ref{fig:QCDflow}.
The propagators in \Fig{fig:QCDflow} are the fully dressed field dependent
propagators and the flow equation describes the fully coupled QCD glue-matter
system.  In particular, there are contributions of the matter sector also in the
diagrams for the gauge sector, i.e., the vacuum polarization diagram in the
gluon propagator, e.g. Fig.~\ref{fig:quark_loop}. The first two loops in
Fig.~\ref{fig:QCDflow} constitute the glue potential of QCD. Note that this is
not the Yang-Mills potential, as the gluon and ghost propagator are those of QCD
and scale differently with momenta. In turn, dropping the matter back-coupling
these two loops correspond to the flow of the glue potential of pure Yang-Mills
theory \cite{Braun:2007bx, Braun:2010cy, Fister:2013bh}, and for more details
see \cite{Pawlowski:2010ht, Haas:2013qwp}. The remaining two loops represent the
flow equation of the dynamical quark-meson model, e.g.~\cite{Schaefer:2004en,
Schaefer:2006ds}.

Following these arguments, we employ the Polyakov--quark-meson (PQM)
model~\cite{Schaefer:2007pw, Herbst:2010rf} as an effective realization for
low-energy QCD. As mentioned in the introduction, the PQM model is an extension
of the quark-meson (QM) model by Polyakov-loop variables which allows to
effectively describe chiral and certain aspects of confinement in QCD. The
Euclidean Lagrangian of this model for $N_f=2$ flavors and $N_c=3$ colors
including a uniform quark chemical potential $\mu$, reads
\begin{eqnarray} \label{eq:pqmmodel}
  {\cal L_\text{PQM}} &=& \bar{q} \left(\dr + h (\sigma + i \gamma_5
  \vec \tau \vec \pi ) + \mu\gamma_0\right) q +\frac 1 2 (\partial_\mu
  \sigma)^2 \nonumber \\[1ex]
  &&  +\ \frac{ 1}{2} (\partial_\mu \vec \pi)^2
  +U(\sigma, \vec \pi )  +{\cal U}(\Phi,\bar\Phi; T_0) \,,
\end{eqnarray}
where $\phi=(\sigma,\vec\pi)$ denotes the $O(4)$-symmetric representation of the
meson fields. The scalar sigma meson and the three pseudoscalar pions are
coupled via a flavor-blind Yukawa coupling $h$ to the quark fields, $q$ and
$\bar q$. The covariant derivative $\dr(\Phi)=\gamma_\mu \partial_\mu-i\,g
\gamma_0 A_0(\Phi)$ couples the Polyakov loop, defined as
\begin{equation}
  \label{eq:defphi} 
  \Phi(\vec x) = \frac{1}{N_c} \left\langle \ensuremath{\tr} {\cal P} 
  \exp \left( i\, g\int_0^{\beta} d\tau A_0(\vec x, \tau )\right)
  \right\rangle\,, 
\end{equation}
to the quark fields. In this work we describe the theory solely in
terms of the Polyakov loop $\Phi$ and suppress the
$A_0$ dependence. Differences between these two formulations will be
discussed elsewhere. Moreover, we add a purely mesonic potential
$U(\sigma,\vec\pi)$ that allows for explicit as well as spontaneous
breaking of chiral symmetry
\begin{eqnarray} 
  \label{eq:pot}
  U(\sigma, \vec \pi ) &=& \frac \lambda 4 (\sigma^2+\vec \pi^2 -v^2)^2
  -c\sigma\ \,.
\end{eqnarray}
The explicit breaking of chiral symmetry is realized by a term linear in the
sigma field. The pure quark-meson (QM) model is obtained from the PQM
model by setting $A_0=0$, i.e. $\Phi,\Phibar=1$ and omitting the glue potential
$\mathcal{U}(\Phi,\Phibar;T_0)$.

In Polyakov-loop extended chiral effective models, the glue potential of full
QCD is usually approximated by an effective Polyakov-loop potential
$\mathcal{U}(\Phi,\Phibar;T_0)\,,$ which is constructed by means of a
Landau-Ginzburg-like ansatz. The arising coefficients are fitted to
lattice results for the pure-glue system, see, e.g. \eq{eq:upoly} below.
The explicitly shown parameter $T_0$ determines the scale of the
deconfinement phase transition. 

In the following we apply a simple polynomial ansatz for the glue
potential as introduced in~\cite{Ratti:2005jh},
\begin{multline}  \label{eq:upoly}
  \frac{\mathcal{U}}{T^{4}} = -\frac{b_2(T;T_0)}{2}\Phi\Phibar 
  - \frac{b_3}{6}\left(\Phi^{3}+\Phibar^{3}\right)+ \frac{b_4}{4}
  \left(\Phi\Phibar\right)^{2}\ ,
\end{multline}
with temperature-dependent coefficient
\begin{equation}
  \label{eq:upolypara}
  b_2(T;T_0) =  a_0 + a_1 \left(\frac{T_0}{T}\right) + a_2
  \left(\frac{T_0}{T}\right)^2 + a_3 \left(\frac{T_0}{T}\right)^3. 
\end{equation}
The parameters of Eqs.~(\ref{eq:upoly}) and (\ref{eq:upolypara}) have been
determined in~\cite{Ratti:2005jh} by a fit to pure Yang-Mills lattice results to
be
\begin{equation}
  a_0 = 6.75\ ,\  a_1= -1.95\ , \ a_2 = 2.625\ ,\ a_3 = -7.44\ 
\end{equation}
and
\begin{equation}
  b_3=0.75\ ,\quad b_4 = 7.5 \ .
\end{equation}
In this construction, however, the backreaction of the matter sector
is completely neglected. This deficiency can be overcome by replacing
the constant parameter $T_0$ by a flavor number, $N_f$, and quark
chemical potential, $\mu$, dependent function: $T_0\rightarrow
T_0(N_f,T,\mu)$.  For the $N_f$ and $\mu$ dependence of $T_0$ we
follow the arguments in \cite{Schaefer:2007pw, Herbst:2010rf}, to wit
\begin{equation}\label{eq:T0mu}
  T_0(N_f,T,\mu) = T_\tau e^{-1/(\alpha_0 b(N_f,T,\mu))}\,,
\end{equation}
with the $\tau$-scale $T_\tau$ and
\begin{equation}\label{eq:bmu}
  b(N_f,T,\mu) = \frac{11 N_c - 2 N_f}{6 \pi} - b_\mu \frac{\mu^2}{(\hat\gamma
  \ T_\tau)^2}\theta_b(T,\mu)\,.
\end{equation}
For details on our choice of parameters see~\cite{Herbst:2010rf}. 
The sensitivity of the phase structure to the parameter $\hat\gamma$ is
discussed in Appendix~\ref{app:gamma}. There, we have varied this
parameter in the range $0.5\leq \hat\gamma\leq\infty$\,. The lowest value,
$\hat\gamma=0.5\,,$ corresponds to an unphysical $T_0(\mu)\,,$ which tends to
zero already at small $\mu\approx300$~MeV. Hence we have chosen
$\hat\gamma=0.85$ for all figures in the main text, which is also suggested by a
comparison to the HDL approximation (see also the discussion in
\cite{Schaefer:2007pw}). With this value, the chiral and deconfinement
transitions lie close to each other throughout the whole phase diagram (see
Sec.~\ref{sec:phase} below).

The $\mu$-dependent correction in \eq{eq:bmu} is proportional to a difference of
(baryonic) Fermi-Dirac distributions
\begin{equation}\label{eq:theta_T}
 \theta_b(T,\mu) = n_b(T,\mu) + n_b(T,-\mu)- 2  n_b(T,0)\,,
\end{equation}
with
\begin{equation}\label{eq:nb}
  n_{b}(T,\mu)= \frac{1}{1+e^{3(m_q-\mu)/T}}\,,
\end{equation}
and $m_q = h\,\sigma_{\rm vac}$ and $\sigma_{\rm vac} = f_\pi=93$~MeV.
The physical importance of this expression is seen in the limit of
vanishing temperature. Then $\theta_b$ reduces to the step function
$\theta(\mu-m_q)$, as it should in order to account for the
Silver-Blaze property of QCD: at vanishing temperature, chemical
potential effects are expected to only set in when $\mu\geq m_q$. Note
that in the present approximation the binding energy of the nucleons
is neglected. As the temperature increases, the sharp behavior of the
step function is smoothed out. 

The origin of $b(N_f,T,\mu)$ is the flow of the fermionic part of the vacuum
polarization, see \Fig{fig:quark_loop}, or more precisely 
\begin{equation}
\left.\partial_{\vec p^2} \partial_t
  \Pi_{A,k}^{\rm ferm}\right|_{p^2=0}\,,
\end{equation} 
in the presence of the Polyakov loop. This term carries the fermionic
part of the QCD $\beta$ function. Its $\mu$ dependence stems from the
Polyakov-loop enhanced quark/antiquark occupation numbers,
\begin{eqnarray}\label{eq:Nq}
  && \hspace{-.8cm} N_q(T,\mu;\Phi,\bar{\Phi}) \\\nonumber 
    & = &\dfrac{1+2\bar{\Phi} 
    e^{(E_q-\mu)/T}+\Phi e^{2(E_q-\mu)/T}}{1+3\bar{\Phi}
    e^{(E_q-\mu)/T}+3\Phi e^{2(E_q-\mu)/T}+e^{3(E_q-\mu)/T}}\,.   
\end{eqnarray}
with $E_q=(k^2+m_q^2)^{1/2}$, which has been used in
\cite{Braun:2009gm,Pawlowski:2010ht}. Note that the occupation number
\eq{eq:Nq} appears in any quark loop with a Polyakov loop or $A_0$
background. In fact, the flow of the fermionic part of the free energy
both in QCD, \cite{Braun:2009gm,Pawlowski:2010ht} and in the
PQM-model, \cite{Skokov:2010wb,Herbst:2010rf} as well as	
\eq{eq:freeenergy}, are proportional to \eq{eq:Nq}. The latter flow is also
central for the present work, e.g. \eq{eq:freeenergy} below.

Here we use \eq{eq:Nq} to estimate the $\mu$ dependence of the QCD $\beta$
function at vanishing cutoff scale, $k=0$: For large temperatures $T\gtrsim
T_c(\mu)$ and hence deconfining Polyakov loops, $\Phi, \bar\Phi\to 1$, the
thermal distribution factor in \eq{eq:Nq} reduces to the standard Fermi-Dirac
distribution for quarks, e.g. \cite{Braun:2008pi}. In this large temperature
limit we also infer that $N_q\to 1/2$. In turn, for temperatures $T\lesssim
T_c(\mu)$ and confining Polyakov loops with $\Phi,\bar\Phi\to 0$, and
$\sigma=\sigma_{\rm vac}$, we arrive at $n_b(T,\mu)$ in \eq{eq:nb}.
Due to the $\Phi,\bar \Phi$ independence of the large temperature limit we
simply use confining Polyakov loops $\Phi,\bar\Phi= 0$ for all temperatures.
This leads to our final expression \eq{eq:theta_T}.

In summary, the nontrivial factor $\theta_b$ in the definition
$T_0(N_f,T,\mu)$ improves the original choice of $b(N_f,\mu)$ in
\cite{Schaefer:2007pw} with respect to the thermodynamical properties
by taking into account the dynamical change of the quark contributions
to the glue sector for finite temperature and chemical potential.

\subsection{Flow equation for the PQM model}
\label{subsec:flow}

For a realistic description of phase transitions one has to include
thermal and quantum fluctuations. In the present work this is done
within the functional renormalization group (Wetterich)
equation~\cite{Wetterich:1992yh}
\begin{equation}
  \partial_t \Gamma_k[\chi] = \frac{1}{2}\text{STr}\left[
  \left(\Gamma_k^{(2)}[\chi] + R_k\right)^{-1}\partial_t R_k \right]\,.
\end{equation}
In this equation $t=\log(k/\Lambda)$ denotes the RG time, $\chi$ represents a
generic field variable and $R_k$ is the RG regulator that implements the idea of
integration over momentum shells. The supertrace STr, involves a trace over
internal (color, flavor and Dirac) spaces as well as a momentum integration.
Furthermore, it accounts for the correct signs of the fermionic and bosonic
contributions.

After integrating out the gluonic degrees of freedom, the flow of the
free energy contains solely the last two loops in Fig.~\ref{fig:QCDflow}, while
the first two lead to the glue potential $\Omega_{\rm glue}(\Phi,\Phibar)$.
Moreover, this procedure results in modifications of the matter sector due to
the coupling to the Polyakov loops. This is accounted for by our initial
conditions for the quark-meson sector. The QCD free energy is thus given by
\begin{eqnarray}\nonumber 
  \Omega_{\rm QCD}(\sigma,\vec \pi,\Phi,\bar\Phi) & = &
  \Omega_{\rm glue}(\Phi,\bar\Phi) + \Omega_{ {\rm matter},\Lambda}(\sigma,\vec
  \pi,\Phi,\bar\Phi)\\[1ex]
  & + & \int_\Lambda^0 dk\,\partial_k \Omega_{ {\rm matter},k}(\sigma,\vec
  \pi,\Phi,\bar\Phi)\,.
  \label{eq:freeenergy}
\end{eqnarray}
Since the gluons have been integrated out, their dynamics is stored in the
Polyakov-loop glue potential $\Omega_{\rm glue}$. Moreover, the full QCD free
energy is necessarily independent of the cutoff, i.e., $\partial_\Lambda
\Omega_{\rm QCD}=0$. Hence, $\Lambda$-dependent terms appear in $\Omega_{\rm
matter,\Lambda}$.  These can be determined from the flow at $\Lambda$,
e.g.~\cite{Litim:1998nf, Berges:2000ew, Pawlowski:2005xe, Gies:2006wv,
Schaefer:2006sr, Braun:2011pp}. From a phenomenological point of view these
contributions correspond to the high-energy part of the vacuum
fluctuations~\cite{Braun:2003ii}.  Finally, the flow equation for the
matter sector of two-flavor QCD in a Polyakov loop or
$A_0$-background reads~\cite{Braun:2009gm,Skokov:2010wb,Herbst:2010rf}
\begin{eqnarray}
  \label{eq:PQMflow}
   & & \hspace{-.3cm}\partial_t \Omega_k = \frac{k^5}{12\pi^2}
  \left\lbrace \frac{1}{E_\sigma}\coth\left(\frac{E_\sigma}{2T}\right) +
\frac{3}{E_\pi}
  \coth\left(\frac{E_\pi}{2T}\right) \right.\nonumber \\[1ex]
  &&  \left. -\frac{4N_cN_f}{E_q}\left[1 - N_q( T,\mu; \Phi,\Phibar) -
  N_{q}(T,-\mu;\bar{\Phi},\Phi)\right] \right\rbrace\,, \nonumber\\
\end{eqnarray} 
with the Polyakov loop enhanced particle numbers $N_q$ defined in \eq{eq:Nq}. 
The quasiparticle energies are given by $E_i = \sqrt{k^2+m_i^2},\
i=q, \pi, \sigma$, and the masses are defined as
\begin{eqnarray}
  \label{eq:quasienergies}
  m_q^2 & = &h^2\phi^2\,,\nonumber\\
  m_\pi^2 & = & 2\Omega_k'\,,\nonumber\\ 
  m_\sigma^2 & = & 2\Omega_k'+4\phi^2\Omega''_k\,.
\end{eqnarray}
In the above expressions, a prime at the potential denotes the
derivative with respect to $\phi^2$.  

The order parameters $\chi_0 = \left(\sigma_0,\Phi_0,\Phibar_0\right)$
for given temperature and chemical potential are determined by the
solution of the corresponding equation of motion (EoM)
\begin{equation}
  \left.\dfrac{\partial\Omega_{k\rightarrow0}}{\partial\sigma}\right|_{\chi_0} =
  \left.\dfrac{\partial\Omega_{k\rightarrow0}}{\partial\Phi}\right|_{\chi_0} =
  \left.\dfrac{\partial\Omega_{k\rightarrow0}}{\partial\Phibar}\right|_{\chi_0}=
  0\ .
  \label{eq:EoM}
\end{equation}
For the numerical solution of the coupled \eqs{eq:EoM}, we utilize a stochastic
technique which is outlined in Appendix~\ref{app:DiffEv}. In comparison to
standard multidimensional root-finding algorithms, such as Newton's method, we
obtained with this technique a much higher numerical accuracy within reasonable
CPU time, which is required in particular for the evaluation of thermodynamic
quantities.

\subsection{Initial condition at vanishing and finite temperatures and
density}\label{sec:IC}

In order to solve the flow equation (\ref{eq:PQMflow}) numerically,
the parameters $\lambda, v^2$ and $c$ of the meson potential
$U(\sigma,\vec\pi)$ in \eq{eq:pot} as well as the Yukawa coupling $h$
have to be specified at the UV scale.  However, the parameters are not
independent and are related to vacuum low-energy observables in the
IR. For example, the explicit chiral symmetry breaking parameter $c$
relates the pion decay constant and the pion mass via $m_\pi^2f_\pi=c$
and can thus be fixed by these observables. The remaining three
parameters at a given UV cutoff are chosen such that specific values
of low-energy observables ($f_\pi, m_\sigma, m_q$) are reproduced in
the IR.  In particular, the physical mass point is characterized by
the values $f_\pi=\sigma_{\rm vac}=93$~MeV, $m_\pi=138$~MeV and $m_q=h
\sigma_{\rm vac} = 297$~MeV. The only insecure and experimentally not
precisely known quantity is the sigma mass which affects the phase
structure, see, e.g.~\cite{Schaefer:2009ui}. To compare with previous
works we choose for the sigma meson mass $m_\sigma=540$~MeV,
cf.~\cite{pdg:2012}.

To achieve these values in the infrared we have fixed the following initial
values at the UV scale $\Lambda=950$~MeV:
\begin{eqnarray}\label{eq:IC}
 \lambda & = & 1.3\,,\\
  v^2 & = & -4.36\cdot 10^6~(\text{MeV})^2\,,\nonumber\\
  c & = & 1.77\cdot 10^6~(\text{MeV})^3\,,\nonumber\\
  h & = & 3.2\,.\nonumber
\end{eqnarray}
Implicitly, \eq{eq:IC} also takes into account the gluonic fluctuations: below
the UV scale $\Lambda$ they effectively decouple, but their contributions above
the UV scale lead to the initial conditions \eq{eq:IC} and the Polyakov-loop
potential. Moreover, for sufficiently large UV scale $\Lambda$, thermal and
chemical potential modifications are negligible as they are suppressed
exponentially with $-\Lambda/T$ for thermal fluctuations and polynomially with
$\mu/\Lambda$ for density-related fluctuations.  For large temperatures and/or
chemical potential, however, the initial conditions in \eq{eq:IC} receive
corrections.

In general, thermal and quantum fluctuations with momenta $k\leq\Lambda$ have to
be taken into account in a temperature and chemical potential dependence of
\eq{eq:IC}. This can be either done by solving the related QCD flow for $k\geq
\Lambda$ or by projecting this flow on the matter sector. 

In our previous work \cite{Herbst:2010rf} we have improved the thermodynamic
observables by including the integrated UV flow in the initial condition of the
free energy,
\begin{equation}
 \Omega_{\rm matter,\Lambda}[\sigma,\vec\pi,\Phi,\Phibar] = U(\sigma,\vec\pi) +
\Omega_{\Lambda}^\infty[\sigma,\vec\pi,\Phi,\Phibar]\,, 
\end{equation}
evaluated on the EoM. Here, $\Omega_{\Lambda}^\infty$ corresponds to the
integration over the flow for the interacting Polyakov-loop system for
scales $k\geq\Lambda$ and fixed parameters. This ensures the $\Lambda$
independence of system at $k=0$, i.e. RG-invariance. A more detailed discussion
this procedure can be found e.g.~in Refs.~\cite{Braun:2003ii, Skokov:2010wb,
Herbst:2010rf}.

In the present work we extend this procedure to the full initial
condition of the effective action. This is implicitly based on the
results for the coupling flows in QCD with partial or full dynamical
hadronisation \cite{Braun:2009gm,Pawlowski:2010ht,BFHP}: the flow of
the QCD coupling parameters directly related to the parameters
\eq{eq:IC} is small, and hence the computation of the thermal and
chemical potential modifications for $k\geq \Lambda$ on the basis of
the $T=0$ values at $\Lambda$ is already a good estimate of the full
modification in QCD. More details on this will be presented elsewhere.

\section{Mass Sensitivity of the Phase Structure}
\label{sec:phase}

For the subsequent analysis of the mass sensitivity of the phase structure and
thermodynamics, we will only vary the explicit chiral symmetry
breaking parameter $c$, which determines the Goldstone-boson mass and
keep all other parameters fixed.

From the order parameters we can deduce the chiral and deconfinement phase
transition lines $T_\chi(\mu)$, $T_d(\mu)$, respectively, in the $(T,\mu)$-phase
diagram. At physical pion masses and small chemical potential, both transitions
are crossovers, entailing that there exists no unique definition of the
transition temperature.
In the following we use the inflection point of the
corresponding order parameter to define a (pseudo-)critical
temperature. In addition, we compare this transition point to the one defined
by one half of the normalized order parameter
\begin{equation}
  \dfrac{\sigma(T_\chi,\mu_\chi)}{\sigma(0,0)} = \dfrac{1}{2}\,. 
  \label{eq:Tchicrit2}
\end{equation}

In order to highlight not only the influence of fluctuations, but also the
direct impact of the Polyakov loop on the phase structure, we compare results
obtained with the PQM model with those of the pure QM model.
Furthermore, all PQM results presented below include the matter
back-coupling $T_0(N_f,T,\mu)$, as introduced in \eq{eq:T0mu}. 

\begin{figure*}[ht!]
  \centering
  \includegraphics[width=.45\textwidth]{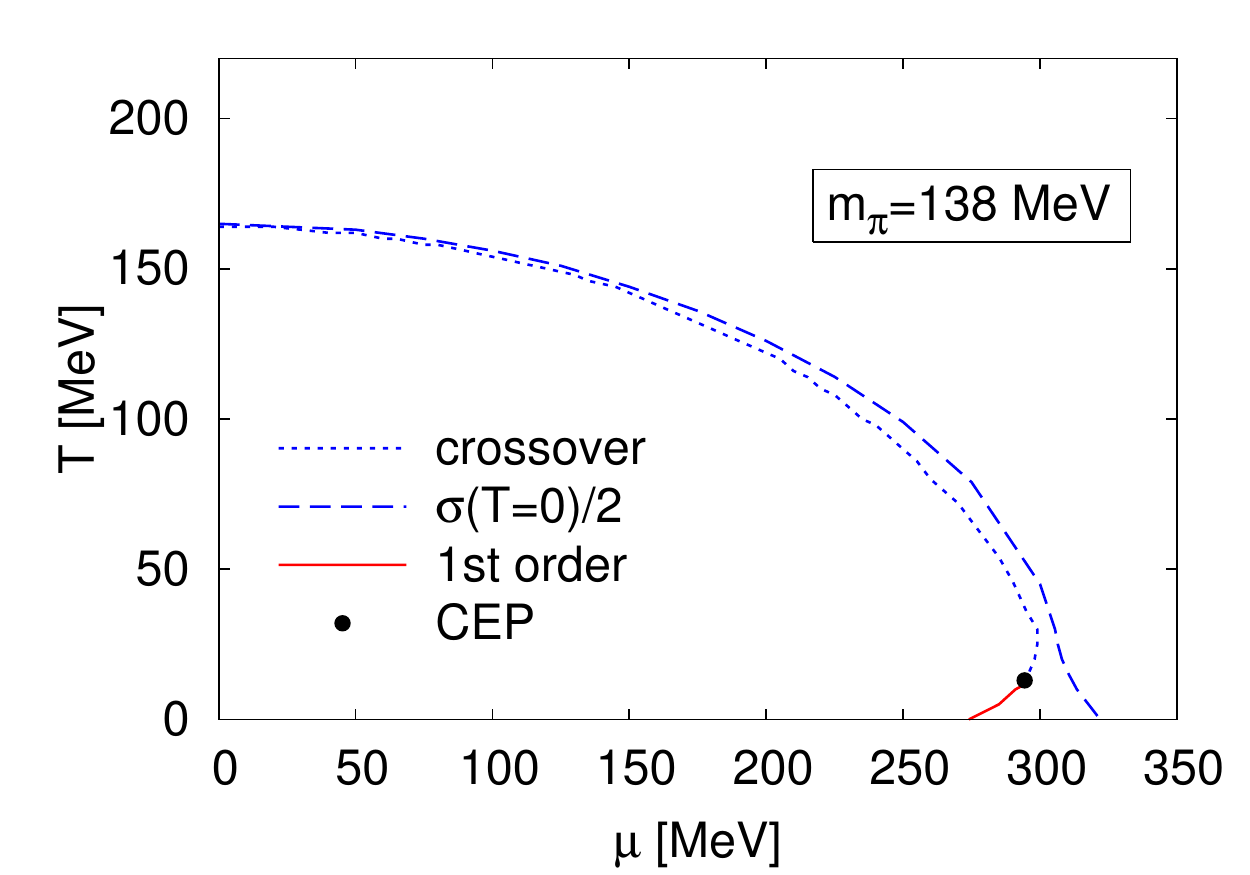}
  \includegraphics[width=.45\textwidth]{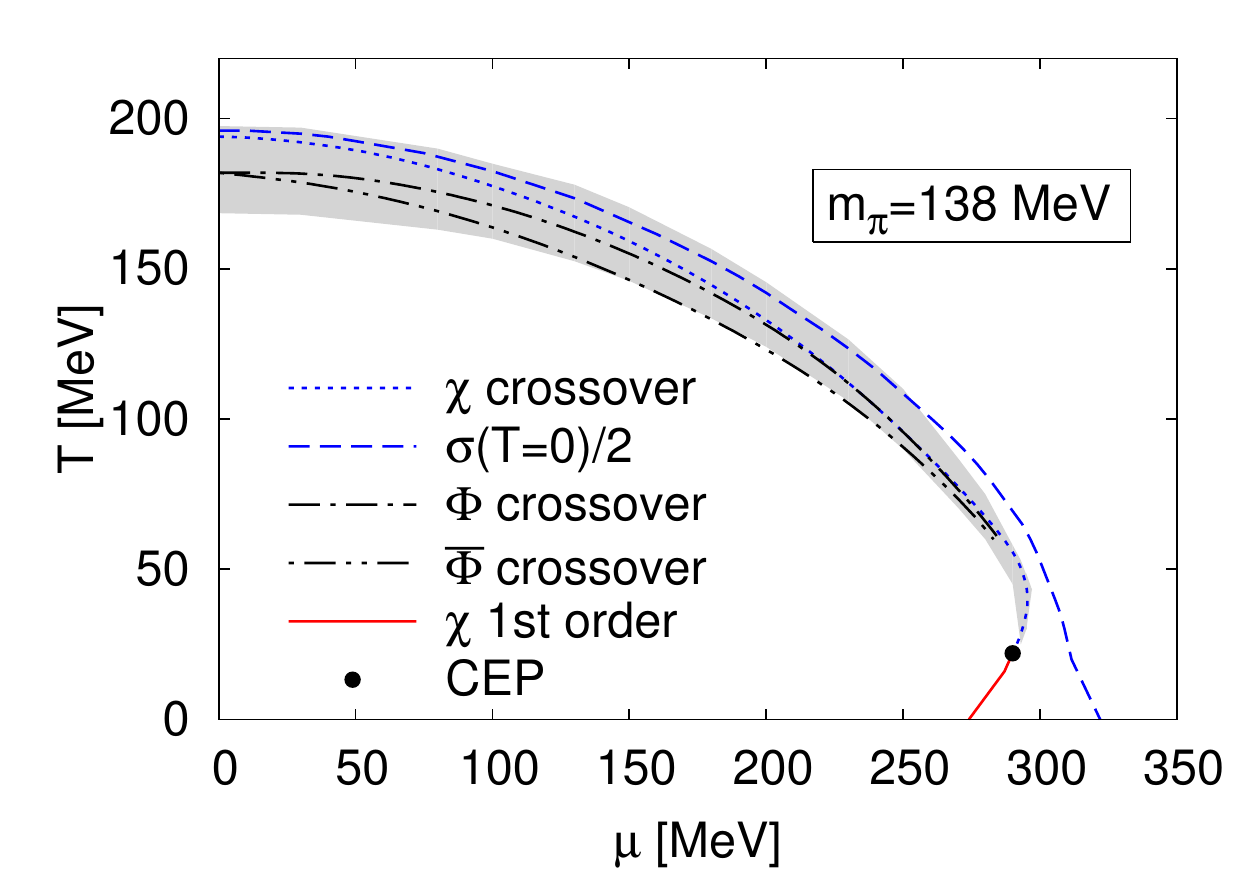}
  \caption{Phase structure at physical pion masses for the QM (left) and
    PQM (right) models. Depicted is the chiral (blue, long- and
    short-dashed) transition line. In the PQM model the Polyakov-loop
    transitions (black, dot-dashed and dot-dot-dashed lines) are also shown.}
  \label{fig:mpi138_phase}
\end{figure*}

\subsection{Physical mass point}

We begin with a discussion of the phase structure for physical values
of the low-energy observables.
The corresponding $(T,\mu)$-phase diagram for the QM model is shown in
the left panel of Fig.~\ref{fig:mpi138_phase}. At low quark chemical
potential we find a chiral crossover around $T \approx 160$~MeV which
turns into a first-order phase transition at large chemical potential
in a critical endpoint (CEP).

In comparison to standard mean-field results, where meson fluctuations are
ignored, we observe that the location of the CEP is shifted towards
lower temperatures when fluctuations are taken into account, see, e.g. also,
\cite{Schaefer:2006ds}. Interestingly, a similar behavior is found if
the standard mean-field approximation is improved by considering the
renormalized QM models. The renormalization of these models amounts to
including vacuum fluctuations of the quark loop
\cite{Nakano:2009ps, Skokov:2010sf}. Already these fluctuations push
the location of the CEP towards higher chemical potential  and smaller
temperatures and exclude the existence of a CEP at low $\mu/T$
ratios. A similar conclusion can be drawn for renormalized (P)QM
models with three quark flavors \cite{Schaefer:2011ex}. Note that these vacuum
fluctuations are always included in the FRG treatment.

In addition, in Fig.~\ref{fig:mpi138_phase} we compare the transition line
obtained by the inflection point of the chiral order parameter (short-dashed
line) with the one obtained via \eq{eq:Tchicrit2} (long-dashed).
At low chemical potential  both curves agree well, but start to deviate at
larger chemical potential . The curve defined by the inflection point runs into
the CEP and the transition turns into a first-order one for smaller
temperatures. A for RG calculations typical back-bending of the transition curve
at low temperatures towards smaller chemical potential  is observed. This
behavior is not seen for the other definition of the chiral transition line,
which does not reach the endpoint. In contrast, the curve bends outwards at
lower temperatures and hits the $\mu$ axis at $\mu\approx322$~MeV.

In the right panel of Fig.~\ref{fig:mpi138_phase} we include the Polyakov loop
and show the PQM phase structure for the same IR values. At vanishing and low
chemical potential, the chiral and deconfinement transitions are both
crossovers (see \cite{Aoki:2006we} for the corresponding lattice results) and
again both definitions of the chiral transition are shown.
For the chiral transition, a similar behavior as in the pure QM model is found.

The Polyakov-loop related transitions, which we refer to as
``deconfinement'' transition lines, are denoted by the two black (dot-dashed and
dot-dot-dashed) lines, defined by the inflection points of the
Polyakov loop and its conjugate. The dark band in Fig.~\ref{fig:mpi138_phase}
denotes the width of $d\Phi/dT$ at $80\%$ of its maximum height and measures the
strength of the phase transition.  At vanishing chemical potential the chiral
and deconfinement transitions lie close to each other and continue to coincide
for increasing chemical potential. However, if the matter back-coupling
is neglected, a splitting of the two transitions at finite chemical potential 
has been observed in previous mean-field~\cite{Schaefer:2007pw, Schaefer:2011ex}
and RG calculations~\cite{Herbst:2010rf}. In particular, the deconfinement
transition line becomes almost $\mu$ independent, i.e., $T_d(\mu)\approx
T_d(\mu=0)$ without the matter backreaction. This effect supported previous
speculations about a possible quarkyonic matter region in the QCD phase diagram,
where chiral symmetry is restored while deconfinement persists
\cite{McLerran:2007qj}. In the present two-flavor study including fluctuations
as well as the matter backreaction we, however, do not observe such a region
anymore. A similar scenario has also been found in recent nonperturbative
Dyson-Schwinger studies for two- and $(2+1)$-flavors, cf. \cite{Fischer:2011mz,
Fischer:2012vc} and references therein. The reader is referred to
Appendix~\ref{app:gamma} for a discussion of the impact of the parameter
$\hat\gamma$ on this conclusion.

Towards the critical endpoint, a narrowing of the dark band is seen that
indicates a sharpening of the transitions.
For temperatures below the CEP, the chiral transition is of first order. Also in
the PQM model, the CEP is located at low temperatures, $(\mu_{\rm CEP}, T_{\rm
CEP})\approx(290, 20)$~MeV.

At this point, a word of caution concerning the high-chemical
potential region should be added. In this region important baryonic
degrees of freedom and diquark fluctuations are usually ignored in the
literature. Due to their complexity this is also done in the present work.
Recent estimates of the influence of baryons on the phase structure for two
colors can be found, e.g., in \cite{Ratti:2004ra, Hands:2006ve, Brauner:2009gu,
Strodthoff:2011tz}.  Due to these omissions the phase structure at high chemical
potential becomes questionable.  However, the region that is to date accessible
to lattice calculations, $\mu/T\leq1$, is well within the region of
applicability of the present study. Hence, we can safely conclude that
for low chemical potential the existence of a critical endpoint can be
excluded, see, e.g., also \cite{Schaefer:2012gy}. 

\subsection{Chiral limit}

In the following, we investigate the mass sensitivity of the phase
structure. As an extreme case we consider the chiral limit, where no
explicit chiral symmetry breaking terms are present and the
Goldstone bosons, the pions, are massless.
\begin{figure*}[ht!]
  \centering
  \includegraphics[width=.45\textwidth]{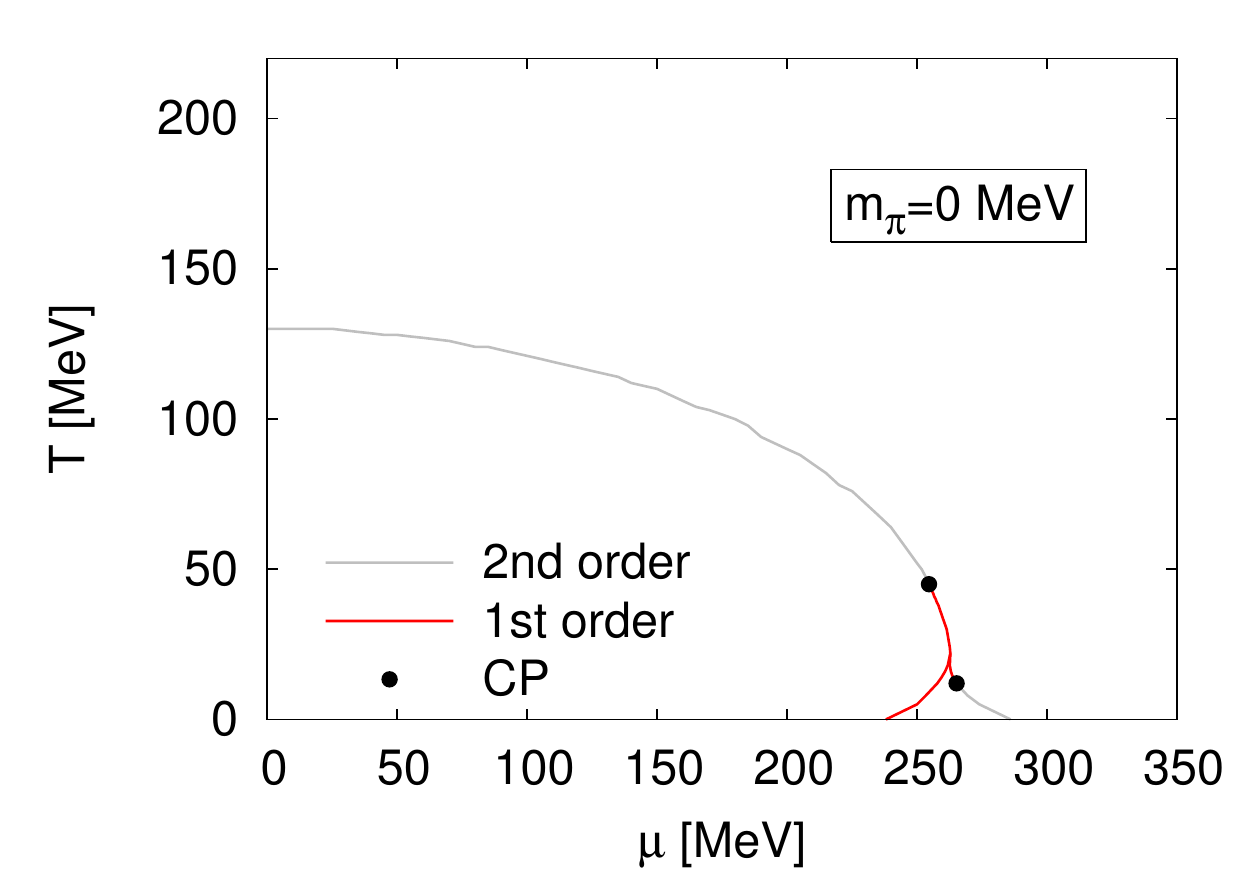}
  \includegraphics[width=.45\textwidth]{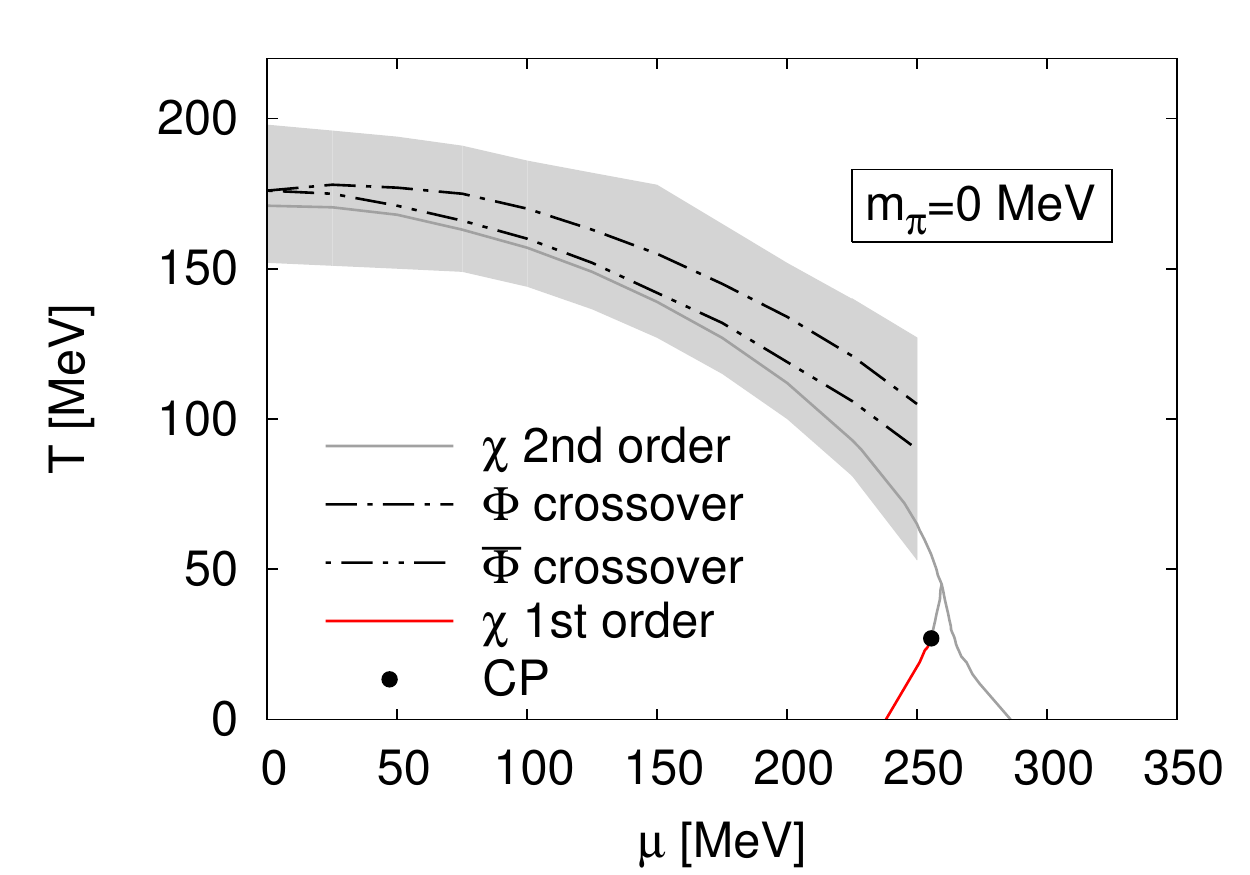}
  \caption{Similar to the previous \Fig{fig:mpi138_phase} but for the
    chiral limit; see text for more details.  }
  \label{fig:mpi0_phase}
\end{figure*}
In this limit the chiral transition at vanishing chemical potential is
a sharp transition of second order lying in the $O(4)$-universality
class, cf.~\cite{Pisarski:1983ms}. The resulting chiral phase structure is
shown in Fig.~\ref{fig:mpi0_phase}. Compared to physical masses the critical
temperature of the chiral transition decreases by approximately
$20$~MeV due to the decreasing constituent quark masses.  Since there
is no ambiguity in the definition of the transition temperature in
this limit, we only show one transition line.

At high chemical potential  and low temperatures, a novel phase
structure emerges: the chiral transition line splits into two
branches. This behavior has previously been observed in a QM model RG
study ~\cite{Schaefer:2004en}; see left panel in Fig.~\ref{fig:mpi0_phase} for
our QM result. In accordance with the previous findings we find two critical
points (CPs) in this region: one on the outer transition branch at low $T$ and a
further CP at higher temperature.

However, for the PQM model only one CP survives which is located on the inner
branch of the transition line, at $(\mu_{\rm CP},T_{\rm CP}) = (255, 27)$~MeV,
similar to the location for physical masses. Below this point, the transition is
of first order, while the outer branch remains of second order. Note that due to
the finite quark masses, the Polyakov-loop related transitions are still
crossovers. Furthermore, the sharp transition in the chiral sector induces an
additional peak in the temperature derivatives of the Polyakov loops. This
interferes with the definition of the dark band in the PQM phase diagram and
subsequently, the focusing of this band towards the critical point is not
directly seen in this limit. 

\begin{figure}
  \centering
  \includegraphics[width=.9\columnwidth]{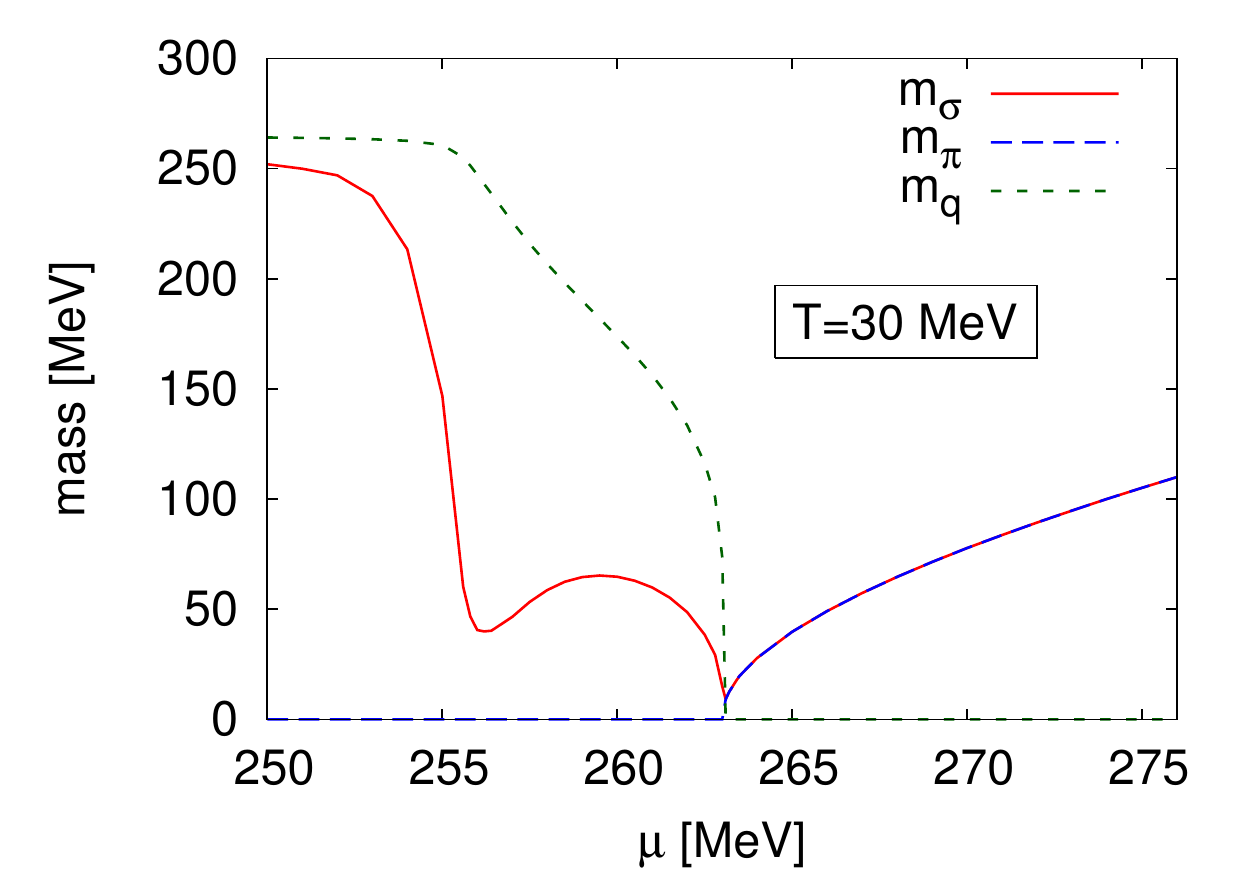}
  \caption{Meson ($m_\pi$, $m_\sigma$) and quark ($m_q$) masses in the
    chiral limit as a function of $\mu$ for constant temperature
    $T=30$~MeV.  One extra minimum of $m_\sigma$ occur in the
    splitting region while $m_\pi$ vanishes until the second chiral
    transition is reached.}
  \label{fig:mpi0_masses}
\end{figure}

Furthermore, the chiral splitting is also reflected in the behavior of
the sigma meson mass as a function of the chemical potential as illustrated in
Fig.~\ref{fig:mpi0_masses} for the PQM model at $T=30$~MeV, i.e. just above the
critical point.  At low temperatures two minima, corresponding to the two
branches of the chiral transition, appear in $m_\sigma(\mu)$. In contrast to the
sigma mass, the pion mass remains zero until chiral symmetry is completely
restored, which happens at the outer transition branch. The quark mass, on the
other hand, is proportional to the chiral order parameter, $m_q = h\sigma$.
Thus it shows a change in slope at the first transition and reaches zero at the
second transition.

\subsection{Small pion mass}

Here we demonstrate how the splitting region in the phase diagram
changes when the pion mass is increased towards the physical point.

\begin{figure*}[ht!]
  \centering
  \includegraphics[width=.45\textwidth]{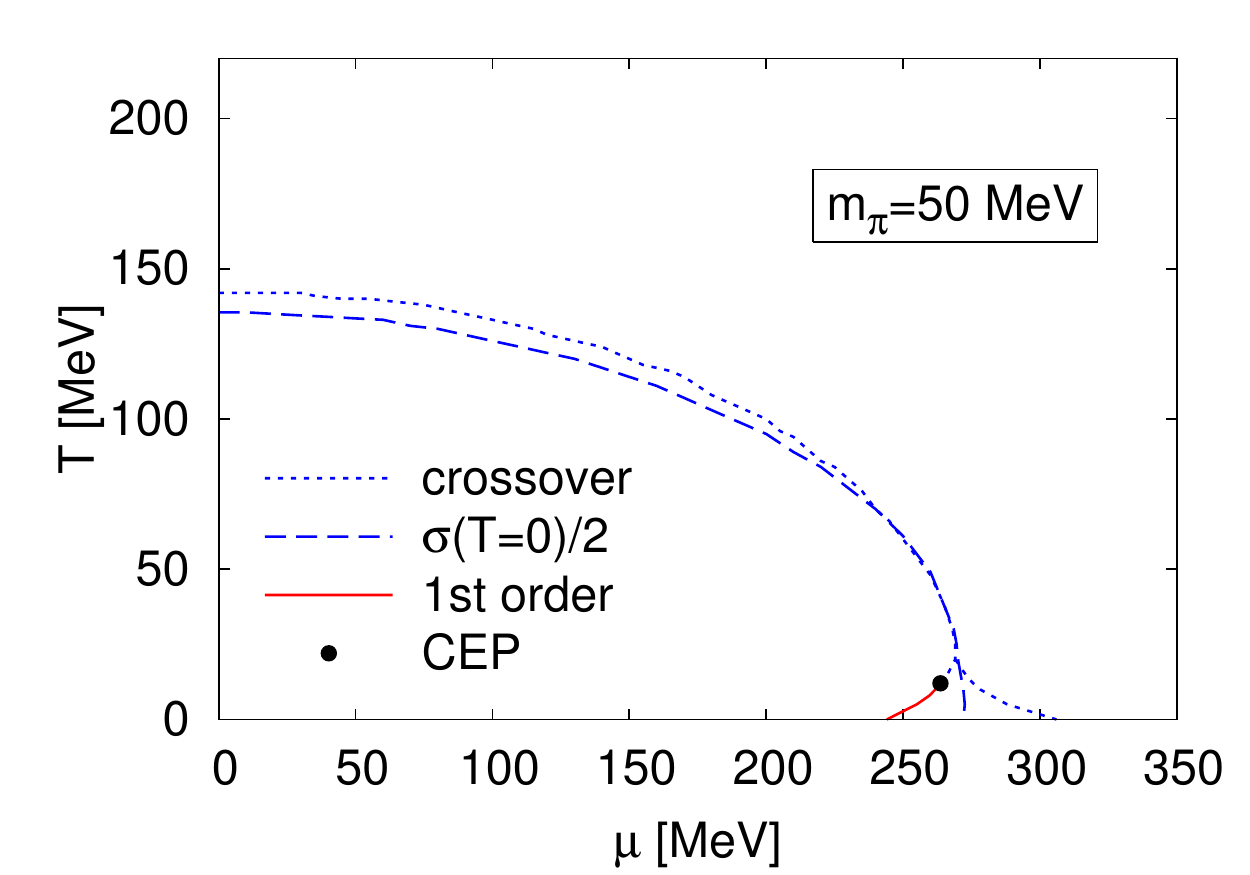}
  \includegraphics[width=.45\textwidth]{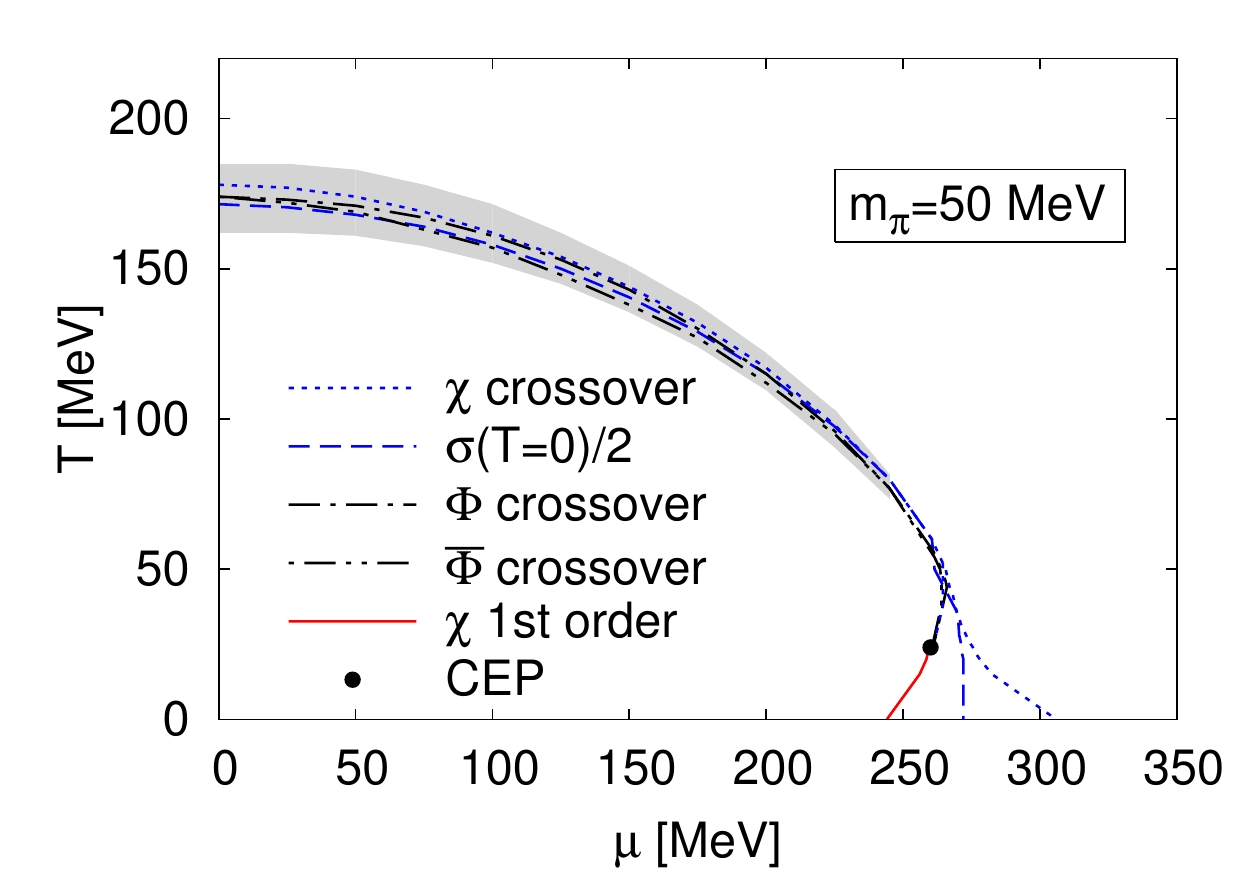}
  \caption{Phase structure of the QM (left) and PQM (right) model for
    small pion masses $m_\pi=50$~MeV.  For these masses, the
    transition splitting persists, while it is washed out for
    increasing masses towards the physical point,
    cf. Fig.~\ref{fig:mpi138_phase}.}
  \label{fig:mpi50_phase}
\end{figure*}

The resulting phase structure for $m_\pi=50$~MeV is shown in
Fig.~\ref{fig:mpi50_phase}. Due to the nonvanishing explicit chiral
symmetry breaking, the chiral transition immediately becomes a
crossover. In the low temperature/high chemical potential region we
still observe a splitting in the chiral transition line and also the
second minimum in the sigma meson mass persists. As the pion mass is
further increased towards its physical value, however, the outer
branch of the transition is weakened.

\begin{figure}
  \centering
  \includegraphics[width=.9\columnwidth]{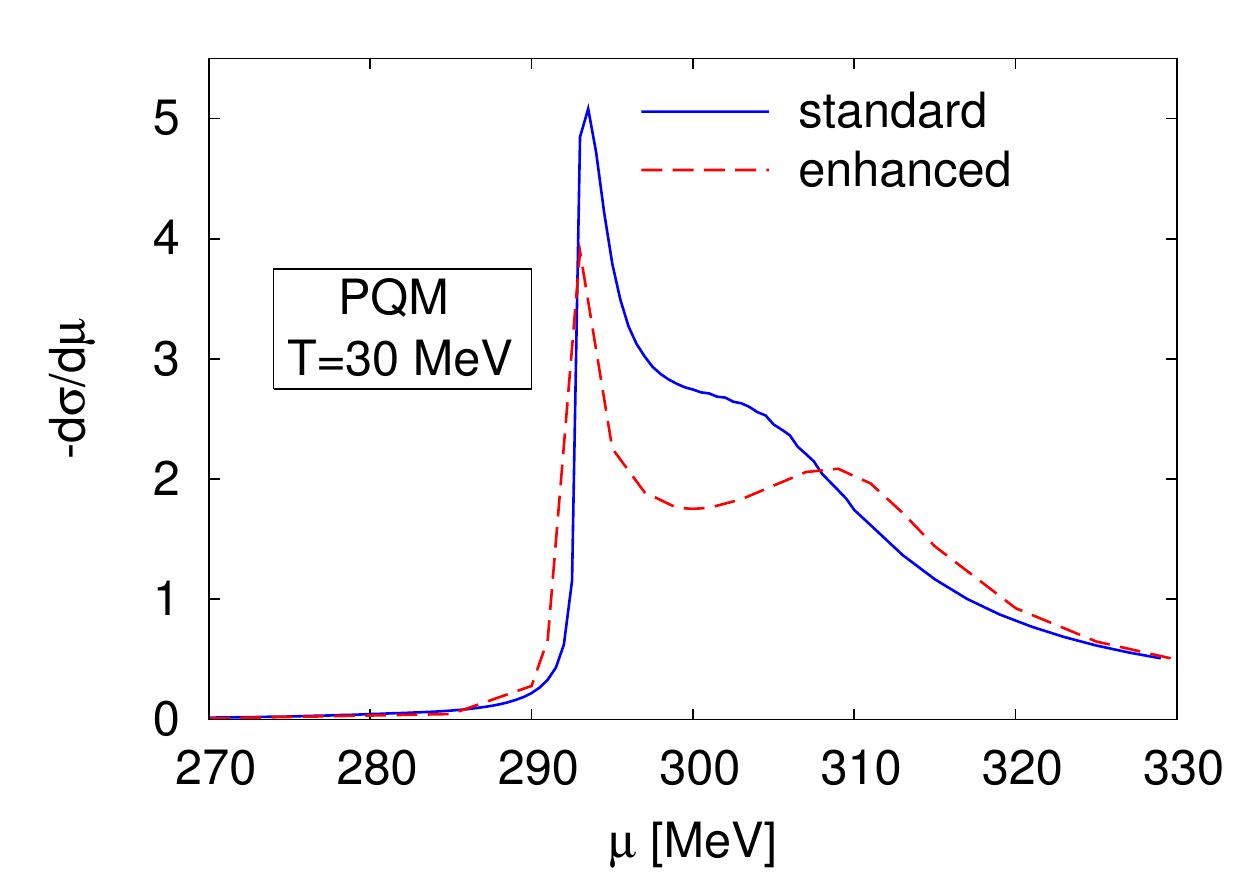}
  \caption{Chemical potential derivative of the chiral order parameter vs
  $\mu$ at physical pion masses and low temperature. Including the $T$ and $\mu$
  dependence of the UV potential, $\Omega_\Lambda(T,\mu)$ as well as the
  modified version of $T_0(\mu)$ a splitting in the chiral transition can be
  determined also at physical masses.}
  \label{fig:mpi138_splitting}
\end{figure}

We also comment on the impact of the thermal and chemical potential
modification of the initial conditions discussed in Sec.~\ref{sec:IC} on the
above effects: when the initial conditions of the flow are fixed in the vacuum
and $T$-independent $T_0(\mu)$ is used, it is not possible to define a second
branch of the chiral transition at the physical point. The inclusion of a $T$
and $\mu$ dependence of the initial action, $\Omega_\Lambda(T,\mu)$, as well as
the more sophisticated definition of $T_0(T,\mu)$ in \eq{eq:T0mu}, however,
enhances this effect once more, as shown in Fig.~\ref{fig:mpi138_splitting}. In
this figure we compare the $\mu$-derivative of the chiral order parameter for
$T_0(\mu)$ and $\Omega_\Lambda(T,\mu)=\Omega_\Lambda(0,0)$ (``standard'') to the
result using $T_0(T, \mu)$ and $\Omega_\Lambda(T,\mu)$ (``enhanced'') at
physical masses. While both versions show a peak at $\mu_\chi\approx293$~MeV --
defining the (inner) transition branch -- the enhanced version shows a clear
additional peak corresponding to the second branch. In fact, the line defined by
this second peak agrees well with the one determined by \eq{eq:Tchicrit2} and is
hence not explicitly shown in \Fig{fig:mpi138_phase}. A comparison
of the phase structure in the standard and enhanced versions can be found
in App.~\ref{app:gamma}.

Note also that the location of the critical point in temperature
direction, $T_{\rm CEP}$, is only mildly sensitive to the pion mass. In all
considered cases we find a critical point at $T_{\rm CEP}\approx
20$--$30$~MeV.  This is in contrast to previous PQM model results where
($T_{\rm CEP}$, $\mu_{\rm CEP}$) depends strongly on $m_\sigma$ when
$m_\pi,\ f_\pi$ and $m_q$ were fixed to their physical values;
see~\cite{Schaefer:2008hk} for mean-field results. When fluctuations
beyond the mean-field approximation are taken into account, this
observation still holds. Here, however, we vary only the explicit
chiral symmetry parameter, which results in different values for
e.g. $f_\pi$ and~$m_q$.

On the other hand, the location of the critical point along the
chemical potential axis, $\mu_{\rm CEP}$, is changed more drastically
as $m_\pi$ is varied.  This can be understood by noting that the
critical chemical potential at vanishing temperature, $\mu_c(T=0)$, is
related to the quark Fermi surface and hence to the quark mass which
increases with $m_\pi$, and so does $\mu_c(T=0)$.

\section{Thermodynamics}
\label{sec:TD}

In order to achieve deeper insights into the nature of the phase
transitions and the impact of the Polyakov loop, we focus in the
following on some thermodynamic observables.
\begin{figure*}
  \centering
  \includegraphics[width=.34\textwidth]{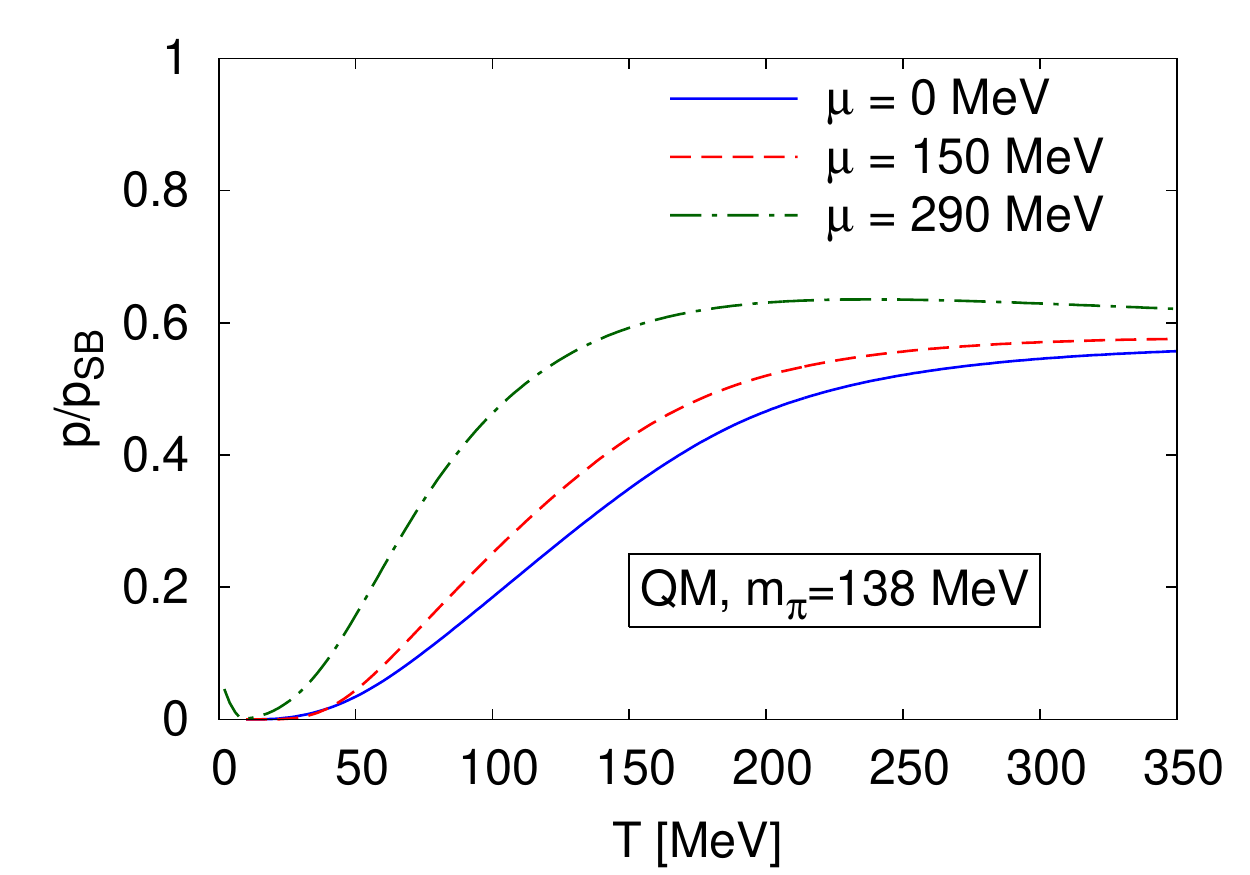}\hspace*{-8pt}
  \includegraphics[width=.34\textwidth]{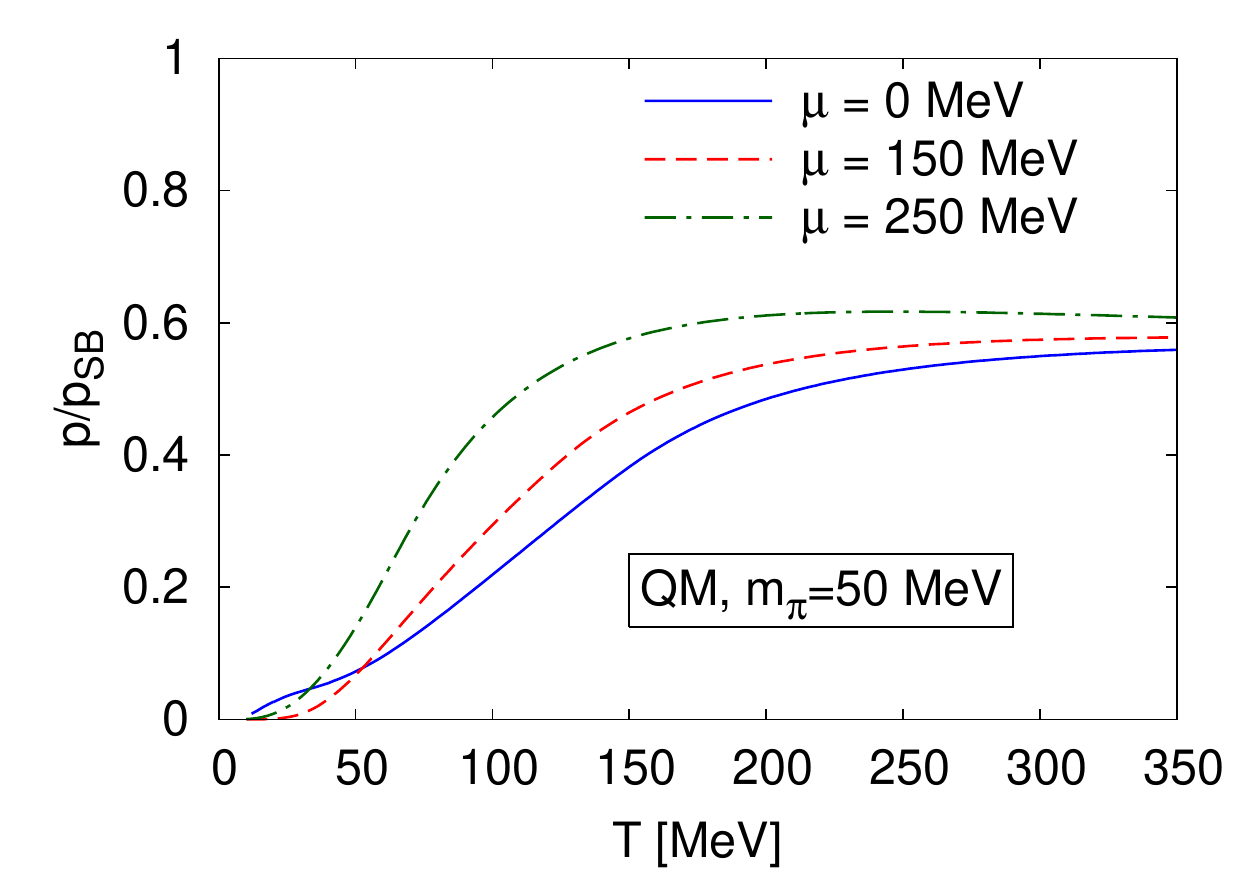}\hspace*{-8pt}
  \includegraphics[width=.34\textwidth]{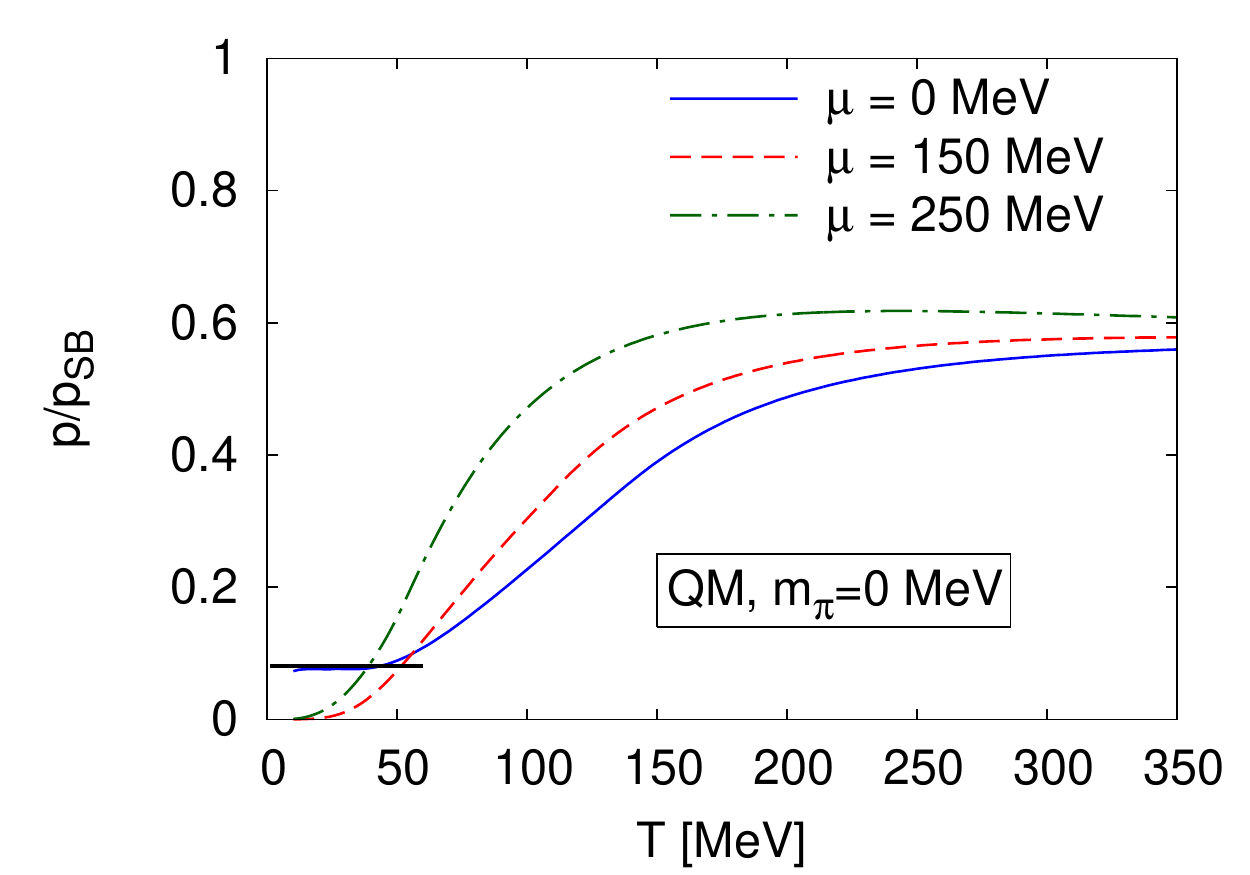}

  \includegraphics[width=.34\textwidth]{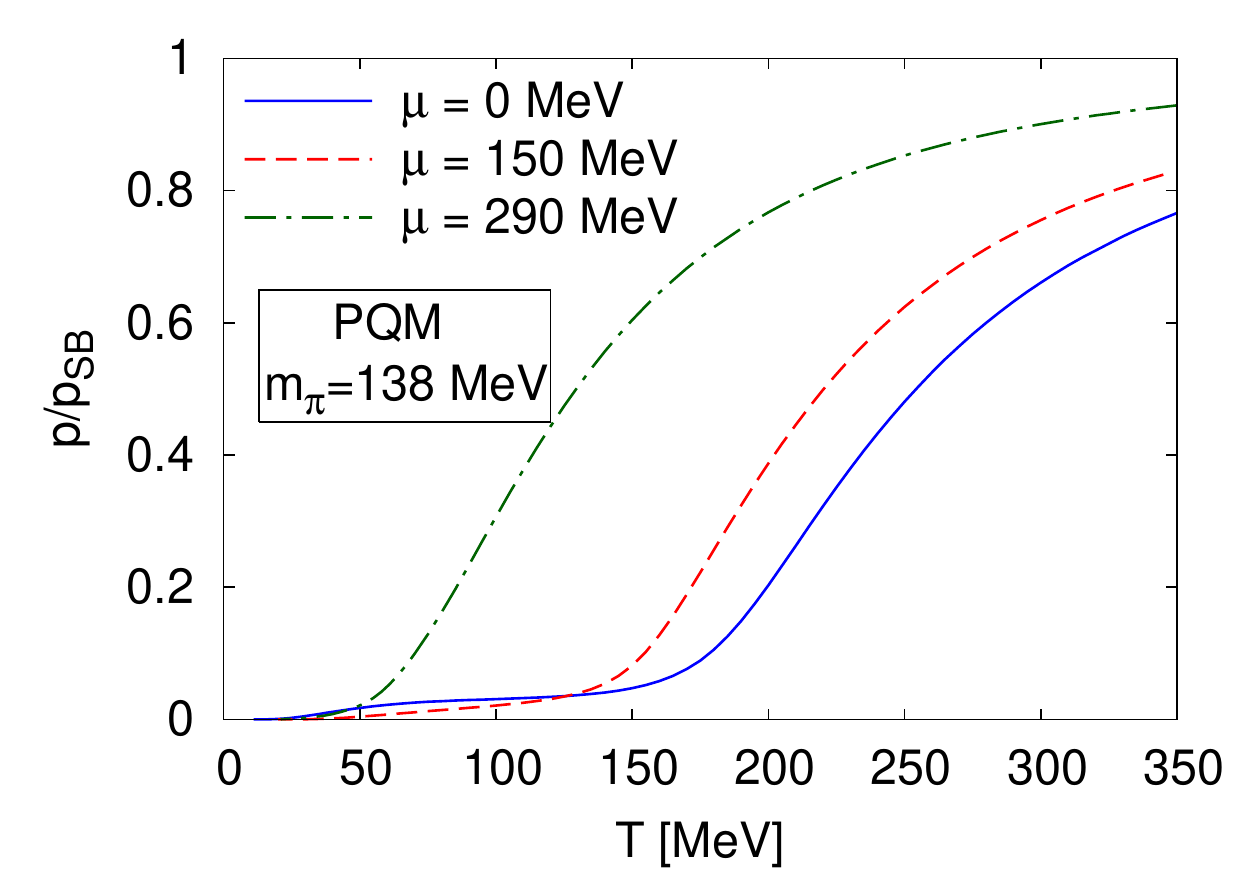}\hspace*{-8pt}
  \includegraphics[width=.34\textwidth]{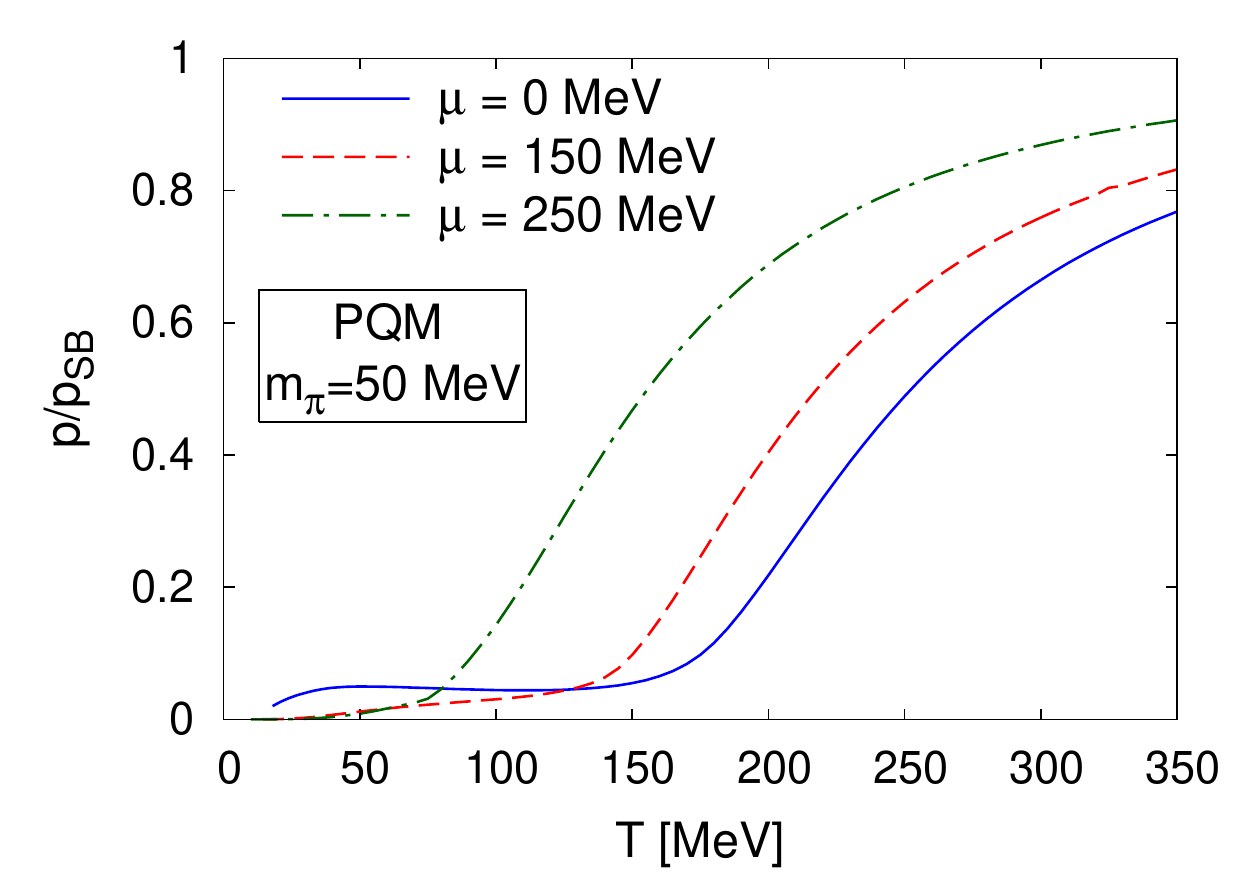}\hspace*{-8pt}
  \includegraphics[width=.34\textwidth]{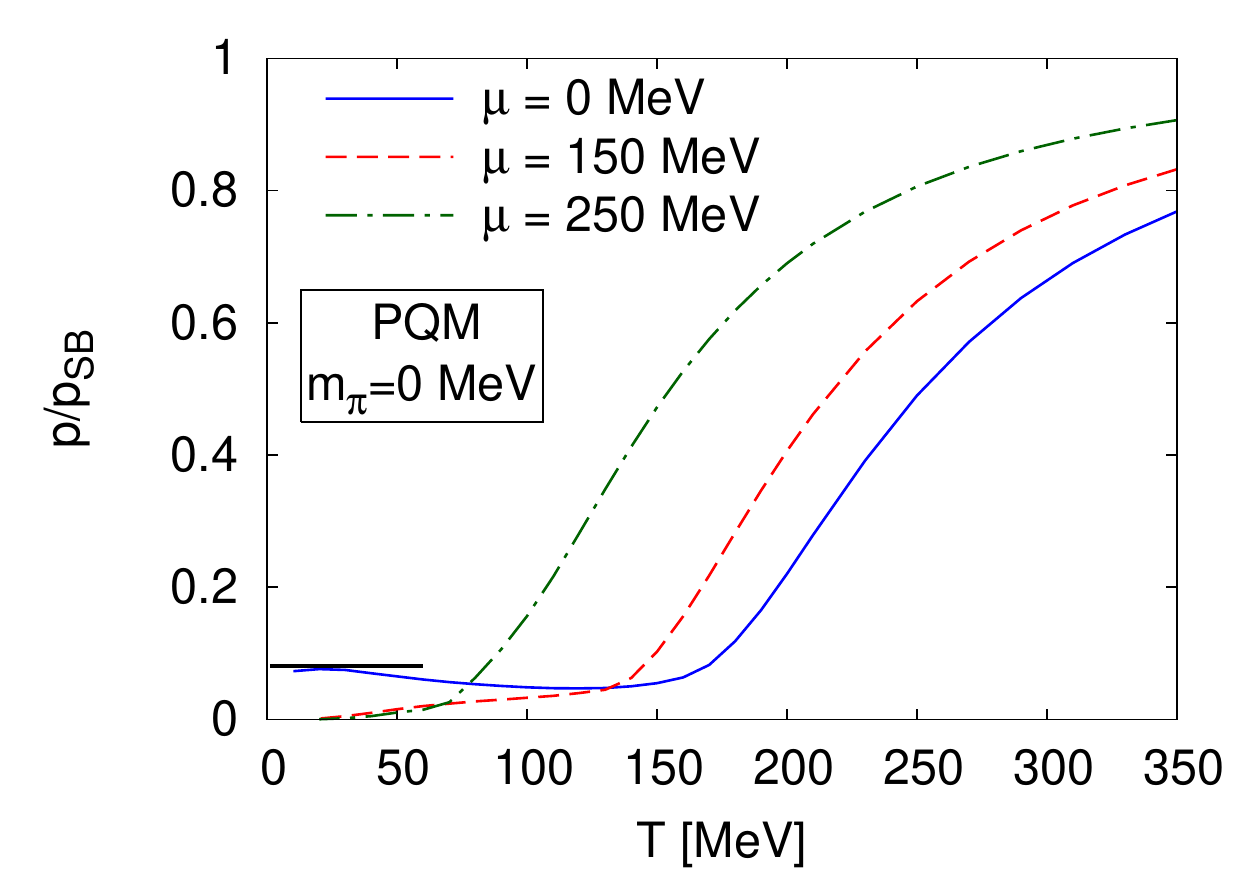}
  \caption{Pressure normalized by the Stefan-Boltzmann pressure,
    \eq{eq:pSB}, in the QM (upper panels) and PQM (lower panels)
    model, each for $m_\pi=138,50,0$~MeV (left to right).}
  \label{fig:QMPQM_p}
\end{figure*}

\subsection{Thermodynamic observables}
The thermodynamic grand potential is obtained by evaluating the infrared
effective average potential on the EoM~(\ref{eq:EoM})
\begin{equation}
  \Omega(T,\mu) =  \left.\Omega_{k\rightarrow0}(T,\mu)\right|_{\chi_0}\,.
\end{equation}
We begin with a discussion of the pressure which is defined as
the negative value of the grand potential
\begin{equation}
  p(T,\mu) = -\Omega(T,\mu) + \Omega(0,0)
\end{equation}
and is normalized to zero in the vacuum. Since the pressure represents a
thermodynamic potential, all further observables follow from this
expression in the standard way by differentiation. For example, the
first derivatives of the pressure with respect to $T$ and $\mu$ yield
the entropy and quark number density
\begin{eqnarray}
  s  =  \frac{\partial p(T,\mu)}{\partial T}\,, &\quad&
  n_q  =  \frac{\partial p(T,\mu)}{\partial \mu}\,,
\end{eqnarray}
respectively. For high temperatures and densities we use the
Stefan-Boltzmann pressure of massless QCD, corresponding to an ideal gas of
massless quarks and gluons for normalization. For $N_f$ flavors and $N_c$
colors
this yields
\begin{eqnarray}
  \frac{p_\textrm{SB}}{T^4} & = & \frac{N_f N_c}{6} \left[\frac{7\pi^2}{30} +
  \left(\frac{\mu}{T} \right)^2 + \frac{1}{2\pi^2} \left(\frac{\mu}{T}
  \right)^4\right] \nonumber\\
  & + & \ (N_c^2 - 1)  \frac{\pi^2 }{45}\,.
  \label{eq:pSB}
\end{eqnarray}
The expression in the first line denotes the fermionic contribution
and the second line contains the gluonic part. Appropriate derivatives of this
expression can be used to normalize other thermodynamic observables.

In addition we can define the energy density $\epsilon= -p + Ts + 2\mu
n_q$ which is used to calculate the trace anomaly
\begin{equation}
  \frac{\Delta}{T^4} = \frac{\Theta_\nu^\nu}{T^4} = \frac{\epsilon- 3p}{T^4}\,, 
  \label{eq:delta}
\end{equation}
which is related to the trace of the energy-momentum tensor
$\Theta^{\mu\nu}$ and vanishes in a scale invariant theory. 
This quantity thus yields a measure for the breaking of conformal
invariance of the system. Furthermore, this observable is also referred
to as the interaction measure, since it quantifies the deviation from the
equation of state of an ideal gas $\epsilon = 3 p$, cf.~\eq{eq:delta}.

\subsection{Mass sensitivity}

In the remaining section we investigate the thermodynamic observables,
the impact of the Polyakov loop on these in hot and dense matter and
study their mass sensitivity by varying the pion masses.

Figure~\ref{fig:QMPQM_p} shows the pressure normalized by its
Stefan-Boltzmann value for three fixed values of chemical potential
and pion masses $m_\pi=138$, $50$, $0$~MeV from left to right. The
largest chemical potential is chosen such that we pass close to the
critical point. Results for the QM model are depicted in the upper
panels, while the PQM calculation is presented in the bottom panels.

The QM pressure levels at $p/p_{\rm SB}\approx 0.6$, owing to the lack
of gluonic degrees of freedom. At low temperatures, the pressure is
dominated by the lightest mesons, i.e. pions. Due to the complete lack
of confinement in the QM truncation, the pressure rises almost
immediately, while the slope becomes steeper for higher chemical
potential. As the pion mass is lowered, the expected dominance of the
Goldstone bosons in $p$ is visible \cite{Schaefer:1999em}. In the
chiral limit and at vanishing chemical potential, a plateau develops
that is slightly below the expected value of $p/T^4=3 \pi^2/90$ for a
free gas of massless pions, denoted by the black line at low $T$.

Including the Polyakov loop, which represents a statistical
implementation of confinement, the picture changes. Now, the pressure
is almost constant in the hadronic phase. Again, a plateau develops at
low temperatures that becomes more and more pronounced as we decrease
$m_\pi$. With increasing temperature, as the quark masses decrease
due to the restoration of chiral symmetry, the pressure rises more
strongly. As the chemical potential increases, the chiral and
deconfinement transition temperatures are lowered and the pressure
rises strongly already at lower temperatures. It saturates at about
$80\%$-$90\%$ of the Stefan-Boltzmann limit at high temperatures. A
similar behavior is found for all pion masses.

When decreasing the pion mass, however, another problem arises: The
coupling of the Polyakov-loop and quark sectors seem to be badly
balanced. As a result, we find that the pressure increases slower than
$T^4$ at $\mu=0$, resulting in a nonmonotonic ratio $p/p_{\rm SB}$. We emphasis
that nevertheless, $p$ itself is a monotonically increasing function of
temperature, as it should be.

\begin{figure*}
  \centering
  \includegraphics[width=.34\textwidth]{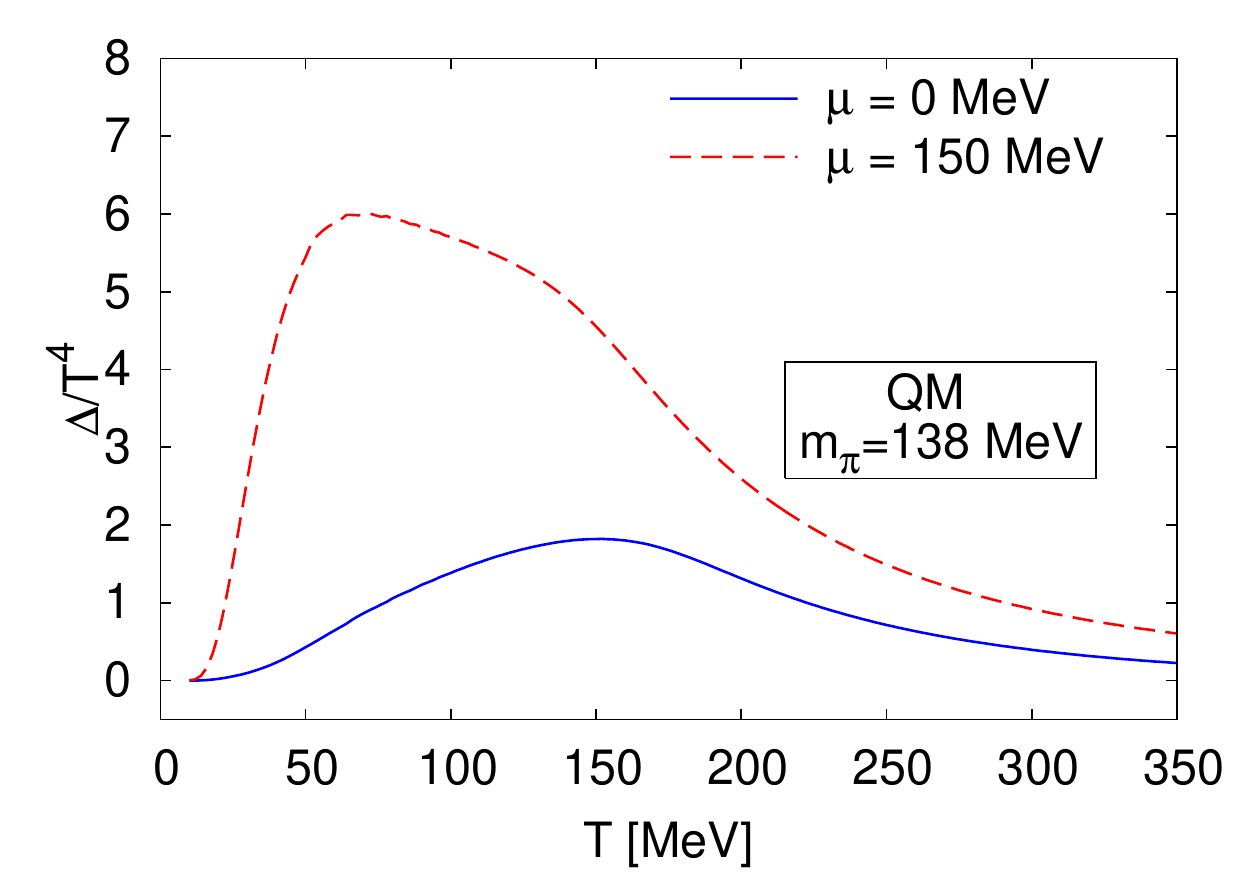}\hspace*{-8pt}
  \includegraphics[width=.34\textwidth]{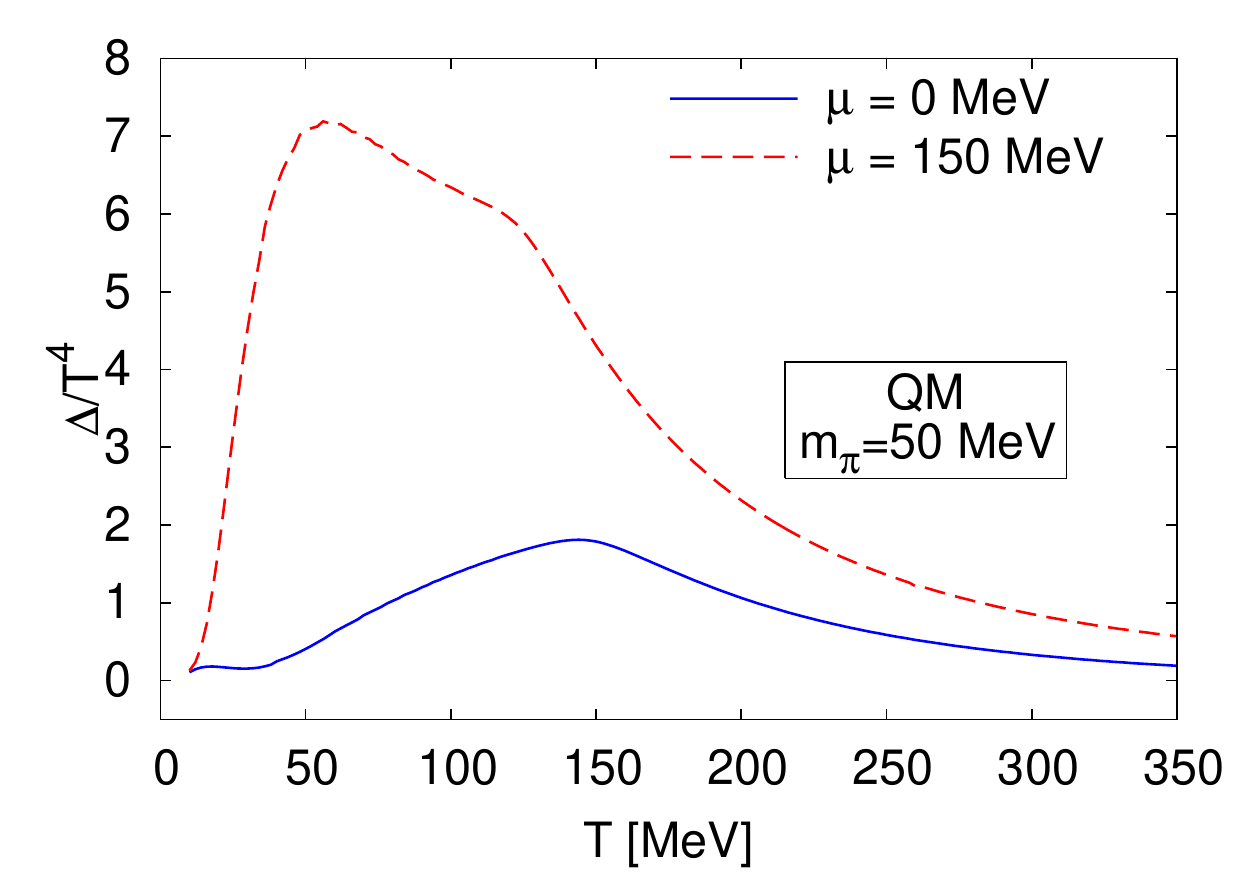}\hspace*{-8pt}
  \includegraphics[width=.34\textwidth]{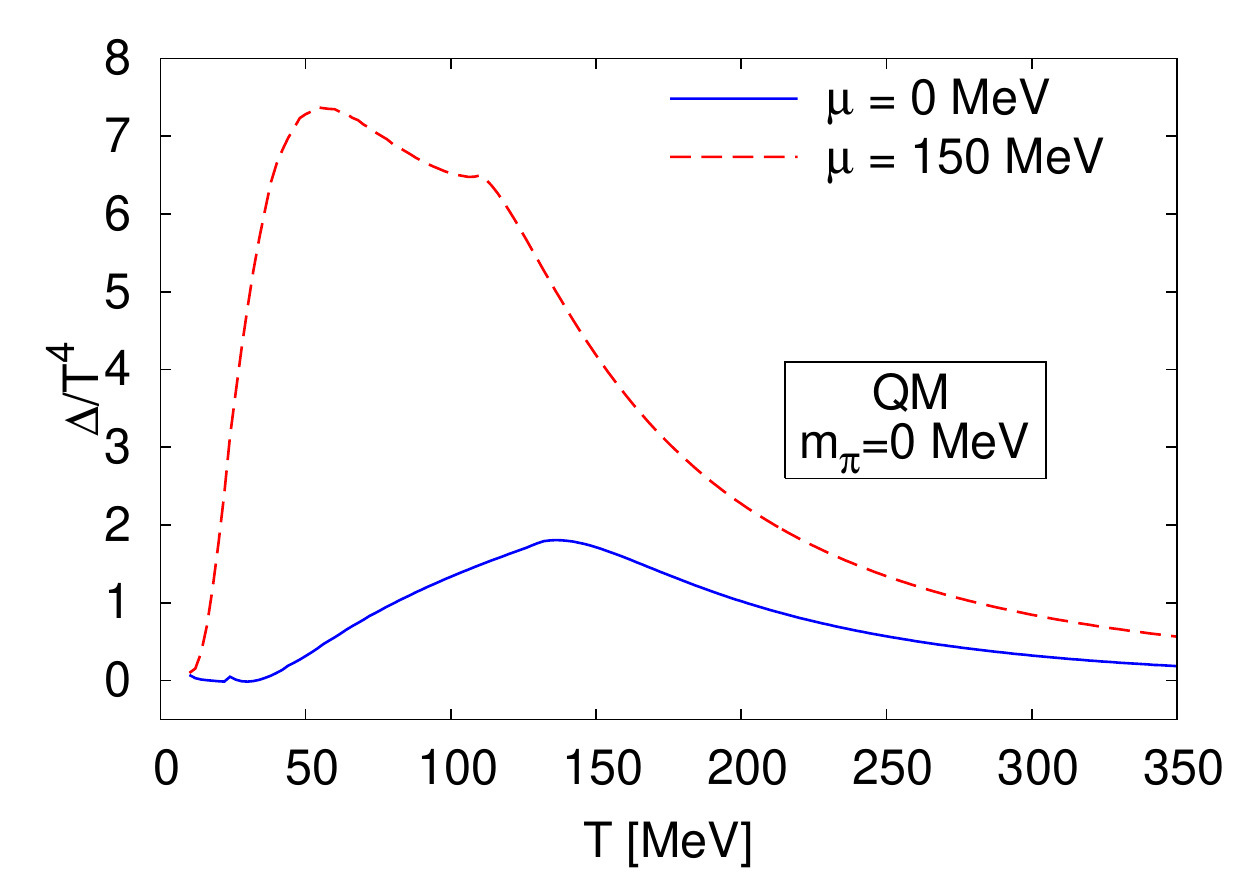}

  \includegraphics[width=.34\textwidth]{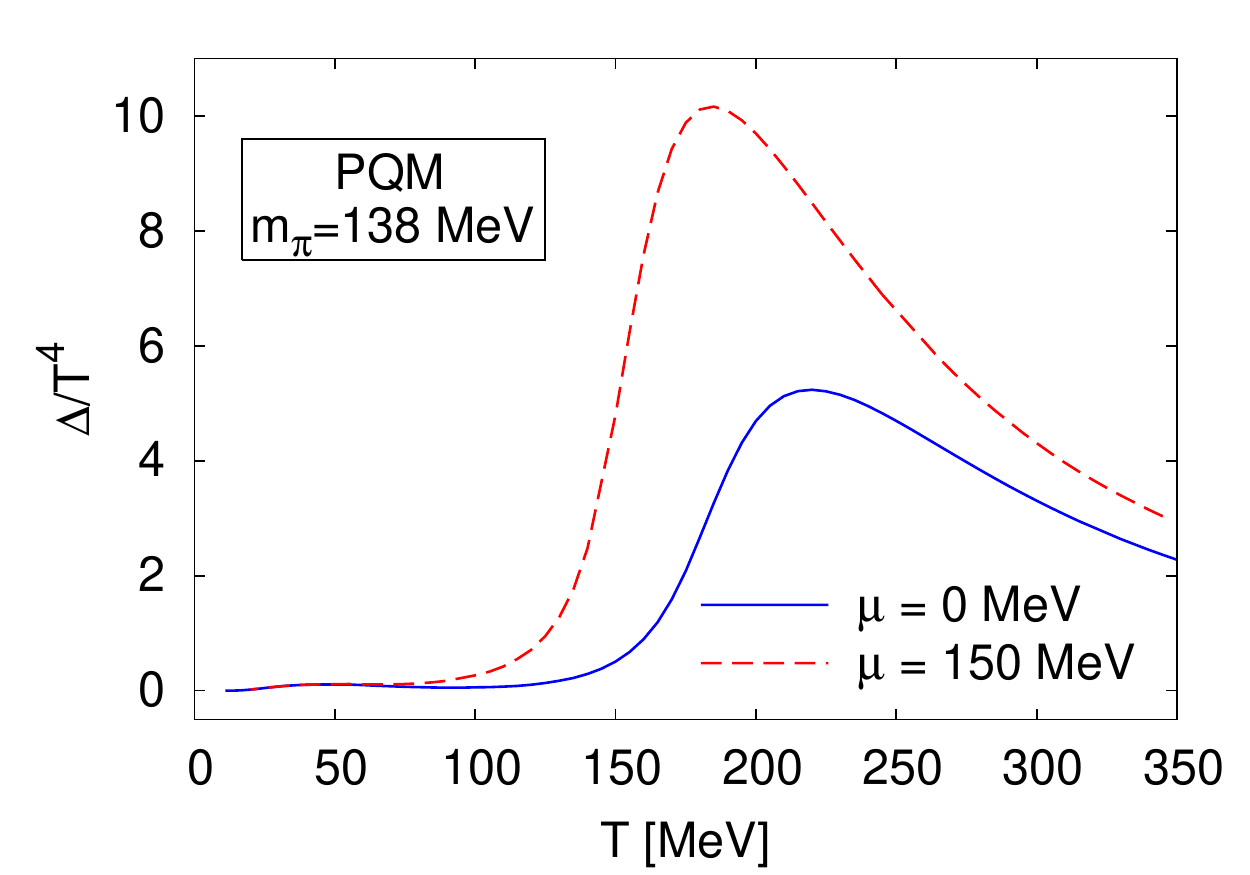}\hspace*{-8pt}
  \includegraphics[width=.34\textwidth]{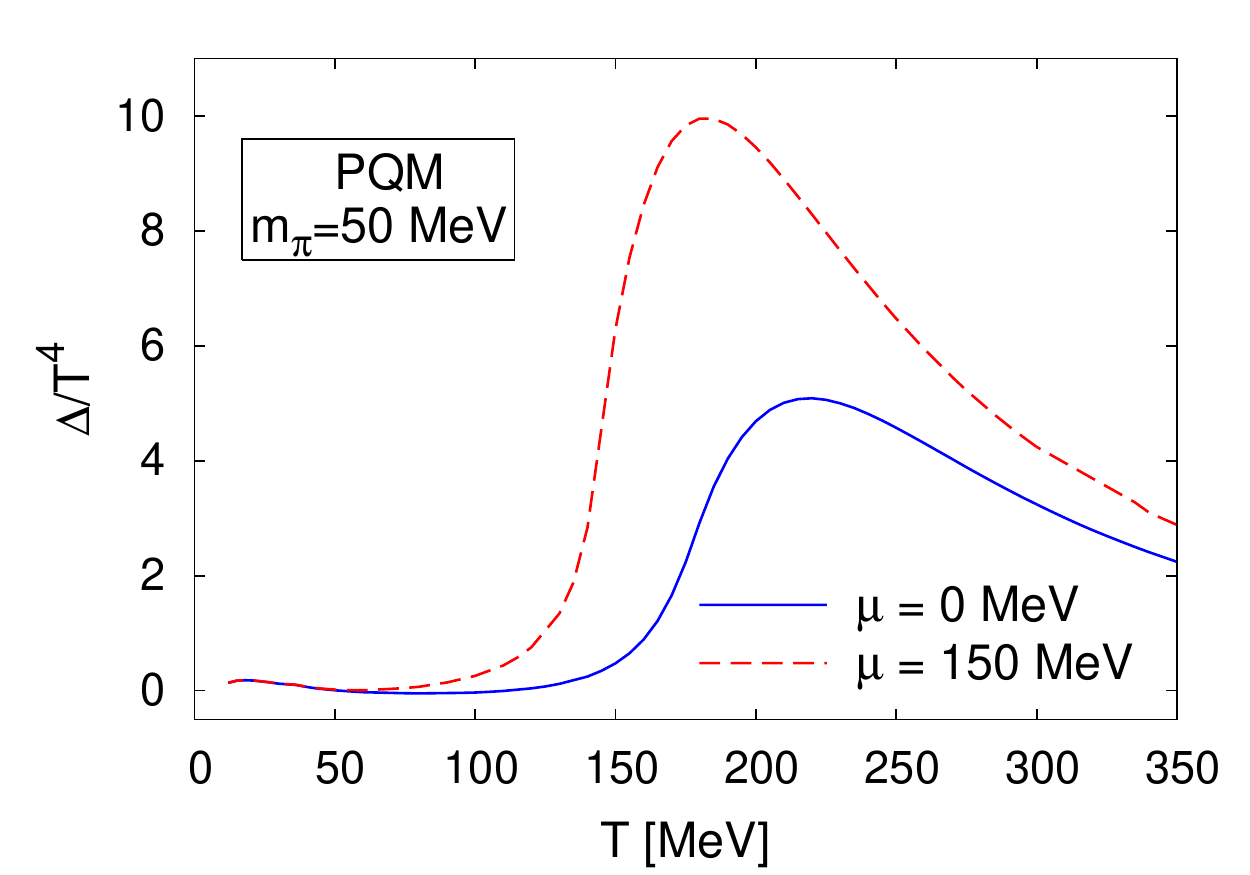}\hspace*{-8pt}
  \includegraphics[width=.34\textwidth]{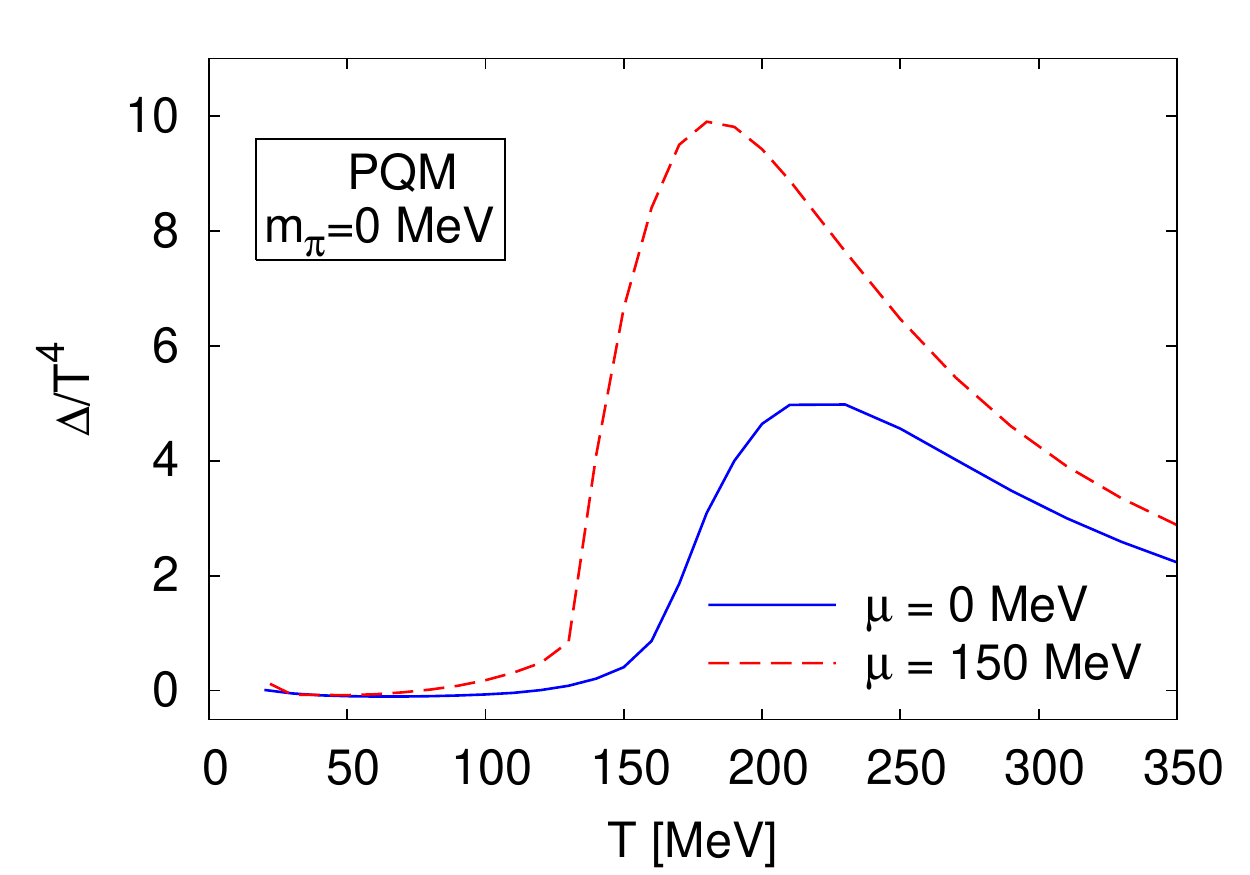}
  \caption{Interaction measure $\Delta/T^4$ in the QM model (upper
    panels) and PQM model (lower panels), see text for details.}
  \label{fig:QMPQM_delta}
\end{figure*}

Combining the energy density and the pressure as defined in \eq{eq:delta}, we
obtain the interaction measure $\Delta/T^4$ which we show in
Fig.~\ref{fig:QMPQM_delta}. This quantity again illustrates
nicely the improvement achieved by the inclusion of the Polyakov loop
\cite{Schaefer:2009ui}. The QM result of this observable showed an
unphysical two-peak structure at nonvanishing chemical
potential. This effect is cured by the inclusion of gluonic degrees of
freedom. The interaction measure is approximately zero at low
temperatures and increases strongly around the phase transition. At
high temperatures it decreases $\sim 1/T^2$. Of course, the
nonmonotonic behavior of $p/T^4$ at $m_\pi<138$~MeV also influences
this quantity, which results in slightly negative scale anomaly at low
$T$ for these masses. For physical masses, however, we find good
agreement with recent lattice results at small $\mu$
\cite{Borsanyi:2012uq, Bazavov:2012bp}.

\section{Conclusions and Outlook}
\label{sec:conclusion}

Based on a renormalization group analysis we have presented results on
the phase structure and thermodynamics of the Polyakov--quark-meson
truncation for two quark flavors, which serves as an effective
model for low-energy two-flavor QCD.  Special emphasis is put on the
influence of thermal and quantum fluctuations, that are crucial for a
proper description of phase transitions. In particular, we have argued
that the PQM model constitutes a well-controlled approximation to
first-principles full QCD. One focus is given to the back-coupling of
quarks to the glue sector of QCD which results in a $N_f$-, $T$- and
$\mu$-dependent modification of the $T_0$ parameter in the
Polyakov-loop potential.

At physical pion masses we find a chiral critical endpoint at much
lower temperatures than found in standard mean-field calculations. This nicely
demonstrates the importance of fluctuations. Moreover, we can rule out the
existence of a critical point for small chemical potential $\mu/T\approx
1$--$2$, see also \cite{Endrodi:2011gv} for corresponding lattice results. 

When studying the mass sensitivity of the chiral phase diagram, we
encounter an intriguing phase structure. In the chiral limit, at high
chemical potential and low temperatures a splitting of the chiral
transition line into two branches is observed. This splitting is not
observed in the mean-field approximation in the same model, even if
the vacuum term of the quark loop is taken into account. A similar
effect has, however, previously been found in a FRG study of the
quark-meson model \cite{Schaefer:2004en}. Differences between the
previous and the present work can be attributed to our higher value of
the UV cutoff and different vacuum parameters. Interestingly, similar
splitting effects also emerge in a finite volume FRG investigation
within a two-flavor quark-meson model, cf.~\cite{Tripolt:2012}.

The splitting of the chiral transition is also reflected in a second
minimum of the sigma meson mass as a function of the chemical
potential as well as in the quark number susceptibility, which are
sensitive to the chiral properties of the system.

Increasing the pion mass towards its physical value we observe that
the splitting in the chiral transition line persists. The outer branch
of the transition is weakened, which can for example be seen in the
behavior of the sigma meson mass. At physical pion mass, the second
transition is mostly washedout. 

Of course, it would be interesting to study the vicinity of the critical point
in more detail, but unfortunately some shortcomings of the Polyakov-loop
potential prevent the computation of thermodynamic observables in the high
chemical potential/low-temperature region. 
The calculation of the phase structure itself relies mainly on the EoM, i.e., on
derivatives of the effective potential with respect to the order parameters.
These are not affected by the above-mentioned problems, which still allows us to
study the phase structure itself.

For a thorough study of the nature of the splitting region at small
pion mass, the deficiency of the Polyakov-loop potential poses serious
problems. Regarding QCD, however, we are aware that in this high
chemical potential region our present approximation is incomplete.  At
baryon chemical potential $\mu_B=3\mu\sim 900$~MeV, baryon degrees of
freedom certainly play an important role in full QCD.  Diquark
fluctuations and baryonic effects are expected to have a sizeable
impact, as also suggested by two-color results~\cite{Ratti:2004ra,
  Hands:2006ve, Brauner:2009gu, Strodthoff:2011tz} and computations
with isospin chemical potential, e.g.~\cite{Kamikado:2012bt}.  Thus,
in order to produce reliable predictions of the physics in the
vicinity of the QCD critical point -- if it actually exists -- an
extension of the present model by the inclusion of these degrees of
freedom is inevitable.

\subsection*{Acknowledgements}
We thank J. Braun, L.~Haas, L.~Fister, M.~Mitter, M.~Puhr,
J.~Schaffner-Bielich, R.~Stiele and M. Wagner for discussions and
collaboration on related topics. This work is supported by the
Helmholtz Alliance HA216/EMMI, by ERC-AdG-290623, by the FWF grant
P24780-N27 and by CompStar, a research networking programme of the
European Science Foundation.  TKH was supported by a
DOC-fFORTE-fellowship of the \"OAW and by the FWF through
DK-W1203-N16.

\begin{appendix}
\section{Differential Evolution Algorithm}
\label{app:DiffEv}
Usually, finding the roots of a higher-dimensional system of nonlinear
equations, such as the EoM, \eq{eq:EoM}, a natural choice would be
Newton's method. This method converges quadratically if the initial
values are already close to the final solution. Applied to our RG
calculation a huge number of iteration steps, of the order
$\mathcal{O}(10^4)$, are actually needed to gain an acceptable
numerical precision of the solution. However, each of these steps
involves a complete RG evolution which increases the CPU time
drastically. Hence, for the computation of thermodynamic quantities
and higher derivatives, in particular for low temperatures, an
alternative method that requires less RG evolutions is
inevitable. Such an alternative technique is provided by the
differential evolution (DE) algorithm for global optimization
\cite{StornPrice1997} which enables the solution of our
three-dimensional system of the nonlinear EoM, \eq{eq:EoM}, to high
numerical precision with fewer RG evolutions.

In detail the system \eq{eq:EoM} is solved as follows: we discretize
the effective potential on a one dimensional grid (in radial direction
$\sigma$) and determine for arbitrary fixed initial values of the remaining
variables ($\Phi,\Phibar$) the potential minimum after the RG evolution. Then
the EoM in $\sigma$ direction is fulfilled and we are left with a
two-dimensional subsystem for the remaining Polyakov-loop variables.

For these variables we define the cost function
\begin{equation}
  f(\Phi,\Phibar) =
    \left(\frac{\partial\Omega_{k\rightarrow0}}{\partial\Phi}\right)^2 
    + \left(\frac{\partial\Omega_{k\rightarrow0}}{\partial\Phibar}\right)^2
    \label{eq:cost}
\end{equation}
and apply the DE algorithm which we describe in the following.
 12.38.Aw 11.10.-z 11.30.Rd 12.38.-t
In the first step of the DE algorithm, an array  $x_{ij}$ of $i=1,\dots,N$ pairs
$(\Phi_i,\Phibar_i)$, called target vector, is generated by randomly choosing
values in a user-specified initial interval. For the present application we use
$(\Phi_i,\Phibar_i)\in[0,1]\times[0,1]$ and typically the target vector has a
length of $N=10-20$.

Step two consists of defining the trial vector $v_{ij}$, which is obtained from
$x_{ij}$ by a randomization procedure. For the details of this procedure, we
refer the reader to \cite{StornPrice1997}.

Subsequently, the cost function is evaluated on the newly defined array
$v_{ij}$ and compared to its value on the current target vector $x_{ij}$. Note
that this is the only step that involves the RG evolution. If, for a given index
$i$, the result of the trial vector is smaller than the one of the target
vector, the corresponding entry in the target vector is replaced by the one from
the trial vector. The overall minimum of the cost function on the modified
target vector can then easily be calculated and compared to the user-specified
criterion, e.g. $f(\Phi_{\rm min},\Phibar_{\rm min})\leq10^{-3}$. If this
criterion
is not fulfilled yet, the procedure is repeated from step two.

With this algorithm it is possible to solve the EoM with an accuracy
of at least $\mathcal{O}(10^{-3})$ after $\mathcal{O}(10^2)$
generations, which corresponds to $\mathcal{O}(10^3)$ RG
evolutions. For comparison, the same number of RG evolutions using
Newton's method would typically yield roots of the order of 
$\mathcal{O}(10^2)$ only.

\section{Parameter Dependence of the Phase Structure}\label{app:gamma}
\begin{figure}
  \includegraphics[width=.9\columnwidth]{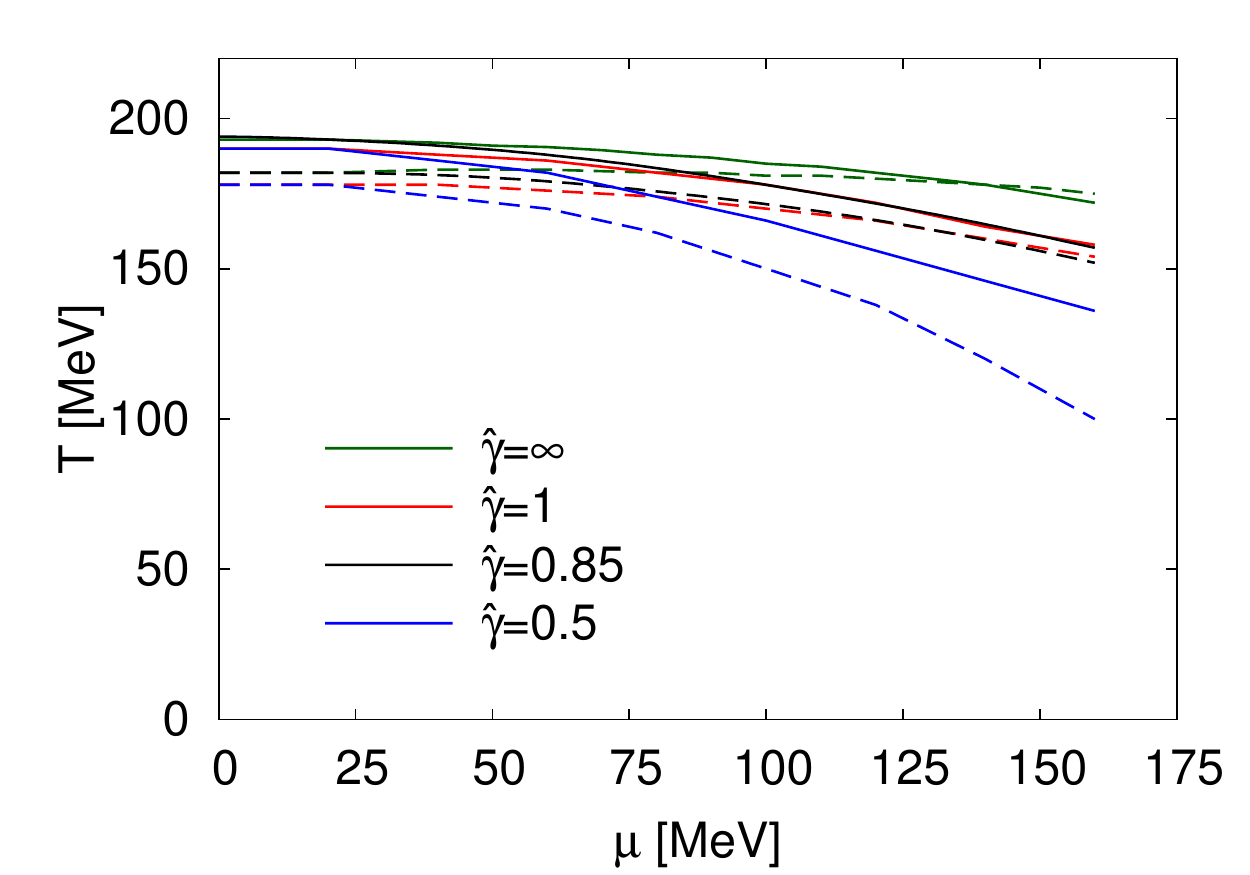}
  \caption{Dependence of the phase structure on our choice of the parameter
    $\hat\gamma$ in $T_0(N_f, T,\mu)$. The solid lines denote the chiral
    transition while the dashed lines correspond to the transition related to
    the Polyakov-loop, $\Phi$. See text for a detailed discussion.}
  \label{fig:gamma}
\end{figure}
\begin{figure}
  \centering
  \includegraphics[width=.9\columnwidth]{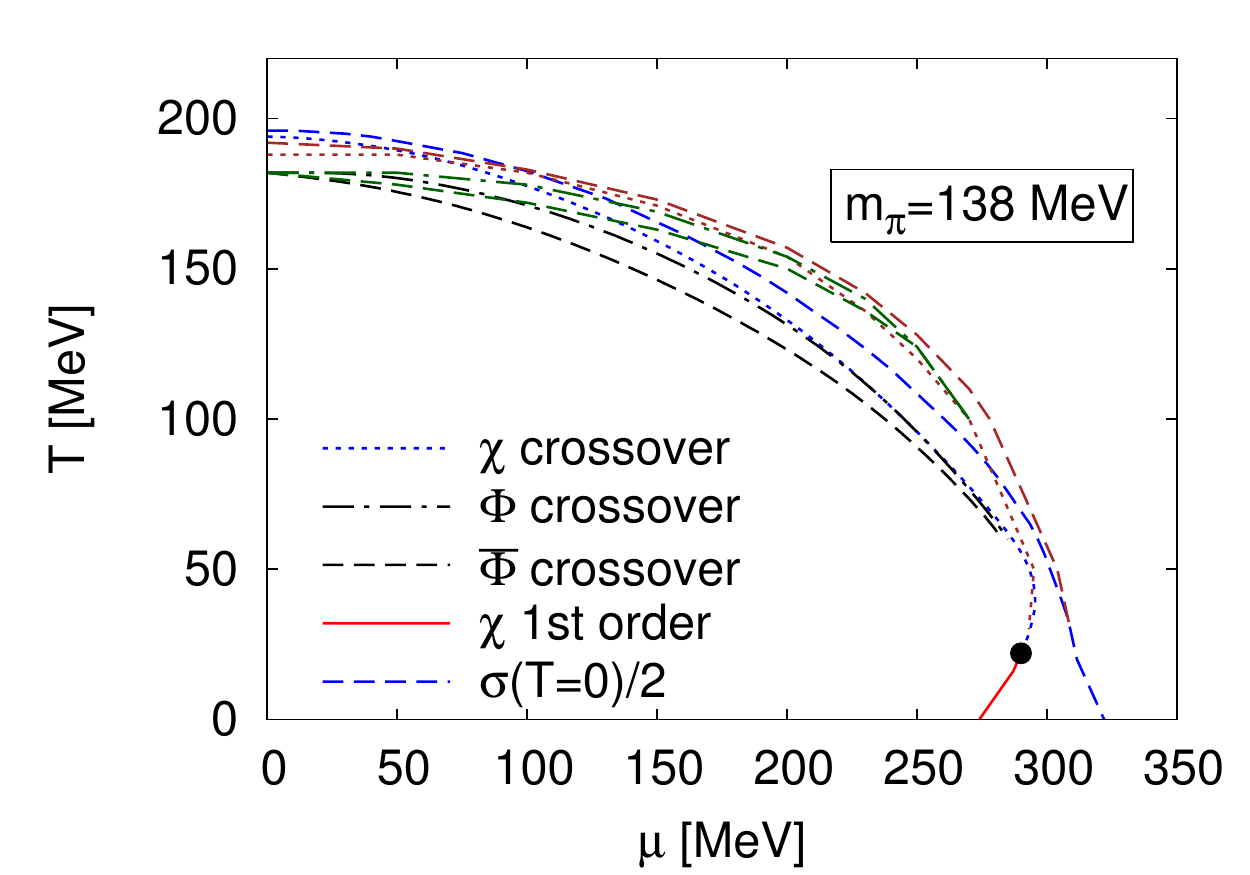}
  \caption{Phase structure at physical masses  with (upper lines) and without
  (lower lines) modified UV potential and $\theta_b$. }
  \label{fig:phase_params}
\end{figure}
In this appendix we briefly comment on the sensitivity of the phase
structure on our choice of parameters for the physical mass point.

First, we consider the parameter $\hat\gamma$ in \eq{eq:bmu}. This quantity
measures the strength of the chemical potential dependence of $T_0$. In
Fig.~\ref{fig:gamma} we show the chiral (solid line) and Polyakov-loop related 
(dashed line) transitions at low and intermediate chemical potentials for
$\hat\gamma=0.5, 0.85, 1, \infty$ (blue, black, red and green lines,
respectively). For better readability we have omitted the line
related to the conjugate Polyakov-loop, $\Phibar$. As can be seen in
Figs.~\ref{fig:mpi138_phase}, \ref{fig:mpi0_phase} and \ref{fig:mpi50_phase},
the corresponding transition always lies below the $\Phi$-related transition
line and both agree within their width. Furthermore, the curvature of both
deconfinement-related transitions is rather similar.

Fig.~\ref{fig:gamma} shows that the curvature of the deconfinement
transition increases continuously as $\hat\gamma$ is lowered from infinity,
which corresponds to constant $T_0$. The full phase diagramm in this case can be
found in Ref.~\cite{Herbst:2010rf}. The curvature of the chiral transition
increases as well, but the effect is much weaker. In the present work we have
used $\hat\gamma=0.85$, which is the value that is suggested by a comparison to
the HDL approximation; see Fig.~\ref{fig:mpi138_phase} for the full phase
structure. For smaller values, the curvature increases even more, see e.g. the
blue curve in Fig.~\ref{fig:gamma}, corresponding to $\hat\gamma=0.5$. Such
values are unrealistic, since they correspond to a $T_0(\mu)$ that approaches
zero already at low chemical potentials.

Next, we investigate the influence of the temperature and chemical
potential dependent initial condition, $\Omega_\Lambda(T,\mu)$, as
well as the temperature dependence of $T_0$, \eq{eq:theta_T}, on the
phase structure. It was already shown in Fig.~\ref{fig:mpi138_splitting} that
the splitting of the chiral transition at high chemical potential becomes
stronger at physical masses when this enhancement is taken into account
(``enhanced version''). In Fig.~\ref{fig:phase_params} we compare the phase
structure for the standard and enhanced versions. The phase structure
with temperature independent $T_0(\mu)$ and fixed initial values
$\Omega_\Lambda(T,\mu) = \Omega_\Lambda(0,0)$ (lower, blue and black
lines) is compared to the enhanced version (upper, brown and green
lines). The results coincide at small chemical potentials. The
curvature of both transitions at low $\mu$ is smaller in the enhanced
version, yielding better agreement with the lattice results. At higher
chemical potentials, the curvature increases once more, bending
towards the result obtained in the standard version at low $T$. As
already pointed out in the discussion of
Fig.~\ref{fig:mpi138_splitting}, in the enhanced version the chiral
transition line defined by \eq{eq:Tchicrit2} agrees with the outer
branch of the chiral transition in this region.

\end{appendix}

\bibliography{../refs}

\end{document}